\documentclass[iicol, pdflatex, sn-basic]{sn-jnl}

\usepackage{multicol} 
\usepackage{graphicx}%
\usepackage{multirow}%
\usepackage{amsmath,amssymb,amsfonts}%
\usepackage{amsthm}%
\usepackage{mathrsfs}%
\usepackage[title]{appendix}%
\usepackage{xcolor}%
\usepackage{textcomp}%
\usepackage{manyfoot}%
\usepackage{booktabs}%
\usepackage{algorithm}%
\usepackage{algorithmicx}%
\usepackage{algpseudocode}%
\usepackage{listings}%
\usepackage{rotating}%
\usepackage{changepage,threeparttable} 
\usepackage{lscape}

\usepackage{hyperref, float}
\hypersetup{
    colorlinks,
    linkcolor={red!50!black},
    citecolor={blue!50!black},
    urlcolor={blue!80!black}
}

\begin{document}

\title[Be and Bn Binary Systems   ]{Exploring the binary origin of B and Be rapid rotators}


\author*[1,2]{\fnm{Jonathan} \sur{Labadie-Bartz}}\email{jbartz@udel.edu}

\author[3]{\fnm{Mark} \sur{Suffak}}\email{msuffak@uwo.ca}

\author[3]{\fnm{Carol} \sur{Jones}}\email{cejones@uwo.ca}
\equalcont{These authors contributed equally to this work.}

\author[4]{\fnm{Ya\"el} \sur{Naz\'e}}\email{ynaze@uliege.be}
\equalcont{These authors contributed equally to this work.}

\author[5]{\fnm{Ken}
\sur{Gayley}}\email{ken.gayley@gmail.com}
\equalcont{These authors contributed equally to this work.}

\author[6]{\fnm{Geraldine} \sur{Peters}}\email{gpeters@usc.edu}
\equalcont{These authors contributed equally to this work.}

\author[3]{\fnm{Rina} \sur{Rast}}\email{krast@uwo.ca}
\equalcont{These authors contributed equally to this work.}

\author[3]{\fnm{Anusha} \sur{Ravikumar}}\email{araviku4@uwo.ca}
\equalcont{These authors contributed equally to this work.}

\author[7]{\fnm{Asif} \sur{ud-Doula}}\email{auu4@psu.edu}
\equalcont{These authors contributed equally to this work.}

\author[2]{\fnm{Coralie}\sur{Neiner}}\email{coralie.neiner@obspm.fr}\equalcont{These authors contributed equally to this work.}

\author[8]{\fnm{Jeremy J.}\sur{Drake}}\email{jeremy.1.drake@lmco.com}
\equalcont{These authors contributed equally to this work.}

\affil[1]{
    \orgdiv{DTU Space}, 
    \orgname{Technical University of Denmark}, 
    \orgaddress{\street{Elektrovej 327}, \city{Kgs., Lyngby}, \postcode{2800}, \country{Denmark}}} 

\affil[2]{
    \orgdiv{LIRA, Paris Observatory}, 
    \orgname{PSL University, CNRS, Sorbonne University, Universit\'e Paris Cit\'e, CY Cergy University}, 
    \orgaddress{\street{5 place Jules Janssen}, \postcode{92195}  \city{Meudon}, \country{France}}}

\affil[3]{\orgdiv{Physics and Astronomy}, \orgname{The University of Western Ontario}, \orgaddress{\street{1151 Richmond Street}, \city{London}, \postcode{N6A 3K7}, \state{Ontario}, \country{Canada}}}

\affil[4]{\orgdiv{STAR Institute}, \orgname{Univ. of Li\`ege}, \orgaddress{\street{All\'ee du 6 Ao\^ut 19c, B\^at B5C}, \city{Li\`ege}, \postcode{B4000}, \country{Belgium}}}

\affil[5]{
    \orgdiv{Department of Physics \& Astronomy}, 
    \orgname{University of Iowa}, 
    \orgaddress{\street{203 Van Allen Hall}, \city{Iowa City}, \postcode{52242}, \state{IA}, \country{USA}}}

\affil[6]{\orgdiv{Department of Physics and Astronomy}, \orgname{University of Southern California}, \orgaddress{\street{825 Bloom Walk, ACB 439}, \city{Los Angeles}, \postcode{90089-0484}, \state{CA}, \country{USA}}}

\affil[7]{\orgdiv{} \orgname{Penn State Scranton}, \orgaddress{\street{120 Ridge View Drive}, \city{Dunmore}, \postcode{PA 18512}, \state{} \country{USA}}}

\affil[8]{\orgdiv{Advanced Technology Center}, \orgname{Lockheed Martin}, \orgaddress{\street{3251 Hanover St}, \city{Palo Alto}, \postcode{94304}, \state{CA}, \country{USA}}}


\abstract{

Observational evidence has continued to mount that a significant fraction of rapidly rotating early-B type stars are products of binary mass transfer. However, very few mid- and late-type B stars with rapid rotation have been demonstrated to be post-interaction products, despite a growing sample of SB1 binaries among stars within this range of spectral types. By considering the currently available information over the entire range of rapidly rotating B-type binaries, we argue that a significant fraction of the mid- and late-type rapid rotators found in binaries are also likely the result of past mass transfer episodes. The observed properties of this sample are compared to the predictions from the Binary Population and Spectral Synthesis code (BPASS), with attention given to the expected evolutionary pathways of stripped stars and the stellar and binary properties of both components of post-interaction systems across a range of initial conditions. 
Prospects for directly detecting and characterizing the stripped cores of the previous mass donors in such systems are described, and the implications for the role of binary interaction in causing rapid rotation are discussed. 
An accurate description of prevalence of binary interaction, the physics of mass transfer, and the post-interaction configuration of systems over a range of initial conditions has far-reaching implications including double-degenerate binaries and their eventual mergers, the output of ionizing UV flux of stellar populations, and the supernova explosions that can arise from stripped or rapidly-rotating progenitors. 
}




\maketitle

\section{Introduction}\label{sec1}

B-emission or Be stars, for short, are B-type stars that produce Balmer emission lines in their spectra due to the presence of a geometrically thin disk of material in the equatorial region of the star, sometimes extending radially out to tens of stellar radii. The disk material is ejected from the star likely due, in part, to rapid stellar rotation (see \citealt{col87} for a commonly quoted definition and the article by \citealt{riv13} for an extensive review of these stars).

These star/disk systems have been studied for over 150 years since they were discovered in 1866 when Angelo Secchi observed the Be star $\gamma$ Cassiopeiae \citep{sec66}, but there remain two major unanswered questions. The first asks what physical process(es), in addition to rotation, act to supply surface material with sufficient angular momentum (AM) to achieve Keplerian orbital velocities leading to the formation of a disk, usually through discrete ``outburst'' episodes. The second relates to the origin of their rapid rotation and high AM content, and how this is maintained through their main sequence (MS) and further evolution. 

The focus of this paper is on the second question, namely the origin of rapid rotation. 
Generally, there are two scenarios that can give rise to the rapid rotation that we observe in Be stars today: such stars may have emerged on the MS already as rapid rotators \citep[e.g.][]{mar06, mar07}, perhaps with evolutionary processes acting to accelerate the surface rotation and keeping in mind that the critical rotational velocity decreases with age as the star expands (\citealt{2008A&A...478..467E, geo13, 2020A&A...633A.165H}) -- this is the single-star evolutionary channel, or they may have acquired an excess of AM during binary mass transfer (the binary interaction evolutionary channel). Evidence continues to mount that a sizable fraction of Be stars were spun up to near-critical via binary mass transfer \citep{van97, has21, wan21}. 
\citet[][and references therein]{jon22} provide a description of theoretical and observational efforts addressing the role of binarity interaction (or lack thereof) leading to rapid rotation, so this is not repeated here. Instead, a brief overview of the key points are discussed below. 
There is no physical reason to suggest the origin of rapid rotation is an ``either/or'' scenario -- both single- and binary-star evolution are capable of leading to rapid rotation. The question remains as to the relative importance of these channels.

There are many reasons why the rapid rotation of these stars, and its origin, is important.  
Rapid rotation impacts the internal structure, not least by making the star non-spherical. It also increases the size of the convective core \citep{nei12} and thus the amplitude of pulsations stochastically-excited in the core \citep{nei20}.
Rotation is believed to enhance mixing of fusion fuel into the core, extending the main sequence and altering stellar evolution, as well as changing surface abundances \citep{mae00}.  
It can also increase mass loss from the most luminous stars \citep{mae00b}, and change the nature of their ultimate supernova \citep{yam05,sum18}.  If rapid rotation is related to binary interaction, as explored further in this paper, it also implies a tight connection between the appearance of rapid rotation, and the birth of massive stars via mass transfer from a companion.  Such rejuvenated stars, and their stripped companions, change in important ways  the ionizing radiation and supernova yields from what would otherwise be expected from a single-star initial mass function (IMF), especially in later times after an initial burst of star formation \citep{got19, got20}.  Hence, understanding the origin of rapid rotation can fill in important details about many aspects of massive-star populations.

Also, this paper focuses on binary mass transfer leading to stripped companions of rapidly rotating B stars.  
These helium-rich stripped companions can be sdOB stars with masses between about 0.3 -- 2 M$_{\odot}$, or high-mass stripped stars between about 2 -- 8 M$_{\odot}$, such as those recently detected by \citet{dro23}.
The somewhat arbitrary distinction is that the latter group can lead to phenomena like core collapse supernovae of hydrogen-free stars and gravitational waves from neutron star mergers, whereas the former can lead instead to binary helium white dwarfs that merge to produce He-rich sdO stars \citep{pan17, dor22} or double-degenerate type Ia supernovae.
Nevertheless, they are both drawn from the same population of stars generated by mass transfer, and their impact on the release of ionizing radiation long after episodes of star formation is related and should be understood together as a unit.

It is important to note that binary mass transfer can result in a rapidly rotating B-type star that does not create a decretion disk because rapid rotation is not a sufficient ingredient to create a Be star.
For simplicity, we will refer to rapidly-rotating B-type stars that have never been observed with emission as Bn\footnote{The ``n” and ``nn” designations date back to the early days of spectral classification, which was done from photographic plates.  Broadened, or ``blurred” photospheric lines from elements heavier than H or He were seen if the $v$sin$i$  of the star was $\geq$ to the plate dispersion in \AA\,/mm \citep{1990clst.book.....J}. ``n” meant  that blurred (nebulous) lines were observed, while ``nn” meant the lines were very nebulous.} stars, and rapid rotators in general (whether or not they posses a disk) as Be/n stars. 

Binarity is important in the upper main sequence, and Be stars are not an exception. However, the observed binary properties of Be stars differ significantly from their non-Be (or more generally, slower rotating) counterparts. For example, there are no known relatively close (P$_{\rm orb} \lesssim 300$ d) early-Be+MS binaries \citep{bod2020}. Instead, nearly all known binary companions to Be stars are evolved. Less attention has been given to the Bn stars in this regard, so the binary properties of the Bn population are poorly understood. It remains an open question if the rapid rotation in Be and Bn stars arises from binary interaction in a similar proportion. 

There are several methods for direct and indirect detection of binarity for Be/n stars. Direct detection of companions typically involves recovering their spectroscopic signature in either UV or visible light (in addition to the rotationally-broadened lines of the Be/n component), or resolving a binary pair with interferometry. Eclipsing binaries (EBs) are also a possibility and detectable through orbital modulation in brightness. In the aforementioned cases, the nature of the companion (whether a stripped star or on the MS) can be determined. In the following cases, the evolutionary status of a companion is not revealed. Radial velocity (RV) measurements of Be/n stars can enable SB1 orbital solutions, providing orbital periods, RV semi-amplitudes of the Be/n star, eccentricity, and a mass function from which the secondary mass can be estimated. For Be stars, their disks can be used to indirectly probe for the presence of a companion. In particular, tidal forces from a companion both truncate the disk \citep{kle19}, and cause spiral density waves \citep[called $m=2$ modes, e.g.][]{pan16, pan18, cyr20, mir23}. While disk truncation may hint at the presence of a companion and indicate an approximate orbital separation, $m=2$ modes can provide an estimate of the orbital period (although the signature of an $m=2$ mode may be observed at $P_{\rm orb}$ or 0.5$\times P_{\rm orb}$). In some cases, when a companion is very hot (as a high temperature stripped star), its radiation field can strongly impact the disk leading to observational features that support the system being a post-interaction binary, such as localized He~I emission with RV motion in phase with the orbit \citep{hum01,kou12,pet13}. 

The aim of this paper is to gather information from the literature about the many individual Be/n stars that are confirmed binaries, and compare this to the MS B+B binary population, while also exploring similarities and differences over the range of Be/n spectral types. To aid in this, the binary population synthesis code BPASS \citep{eld17} is employed to recover the range of initial binary configurations that could result in the sort of post-interaction Be/n binaries that are known, and in particular to make inferences about where low-luminosity stripped stars may be found in the Hertzsprung-Russell diagram (HRD) to guide observational efforts to detect and characterize such objects. Emphasis is given to systems that may have interacted in the past, or may interact in the future -- that is, very wide binaries are not in general considered. 

Section~\ref{sec:overview_bin} provides an overview of the various types of relevant systems in a pre-interaction, ongoing-interaction, and post-interaction configuration. Section~\ref{sec:compare_pops} then examines stellar and binary information from known or candidate Be/n binaries, and compares these to the MS B+B population while also giving consideration to the range of properties within the binary Be/n sample. Section~\ref{sec:BPASS} describes the Be/n binaries (including any known information about the companion star) on the HRD, along with BPASS evolutionary tracks of stripped stars, and Section~\ref{sec:outlook} provides an outlook for observational strategies that can fill in the predicted parameter space that stripped stars are expected to occupy. Conclusions are given in Section~\ref{sec:conclusions}.

\section{Overview of current knowledge of Be/n binaries} \label{sec:overview_bin}

When considering the role of binarity in causing a star to spin up to near critical, the discussion should be focused on relatively close binary systems that can physically interact at some point in the MS or post-MS phase via mass transfer or a merger. Here, we focus only on these cases. There are, however, many examples of Be stars with very wide companions, such as the well-known system $\beta$ Cephei, with a magnetic slowly-rotating pulsating B2III primary, and a mid-type (B5Ve) Be star on an $\sim$81 year orbit \citep{pig92,whe09}. These two stars have not and will not interact, so that the primary plays no direct role in shaping the rotational properties or evolution of the Be star; therefore, we exclude such systems from this study.

In a broad sense, there are two classes of (close) Be/n binaries -- pre-interaction, and post-interaction. There may also be currently interacting systems with accretion disks, but these should be comparatively rare due to the short timescales involved.  There can also be genuinely single Be/n stars\footnote{However, see Section~\ref{sec:runaways} for a brief overview of runaway Be stars, which are presently single but widely believed to be post-interaction systems.}, but these are difficult to confirm as it is challenging to rule out the possibility of faint low-mass companions. The following subsections describe what is currently known about these classes.

\subsection{Pre-interaction binaries} \label{sec:preinteraction}

Possible evidence for single-star or pre-interaction evolution into a Be phase would come from relatively short-period Be+MS systems, but these are apparently extremely rare.
In these cases, it may be presumed that the Be star has evolved as single.
Presently, the two shortest period Be+MS systems are $\alpha$ Eri (Achernar, B6) with an early-A secondary in an eccentric (e = 0.73) seven-year orbit \citep{ker2022} and $\delta$ Sco (B0) in a highly eccentric (e=0.94) 11 year orbit with a main-sequence B2 companion \citep{tan09, tyc11}. 
However, it should be noted that the presence of a main-sequence companion at $\sim$10 AU separation (as in $\alpha$ Eri and $\delta$ Sco) does not necessarily preclude binary interaction in a tight inner binary with separation $<$1 AU.  For example, HD 150136 exhibits a massive inner binary with separation $\sim$0.1 AU, and a slightly lower mass tertiary companion at 12 AU \citep{san13}, which could in time lead to a rapid rotator coupled to a main-sequence star at larger distance as seen in $\alpha$ Eri and  $\delta$ Sco.

It is notable that $\alpha$ Eri and  $\delta$ Sco are the two brightest Be stars in the sky, leaving open the possibility that many more such systems could exist. However, detecting the MS components of $\alpha$ Eri and  $\delta$ Sco is intrinsically difficult without the use of interferometry. 
The radial velocity (RV) variations of the Be star due to binary motion is very small (typically a few km s$^{-1}$), and owing to the long periods and wide Be absorption lines (broadened by their rapid rotation), long time series of high-SNR spectroscopy are required to detect and characterize such orbits.
However, interferometric surveys \citep[e.g.][]{kle24} have recently expanded the limited distances to which main-sequence companions could be detected, yet failed to discover new Be+MS binaries.  This is the basis for regarding such binary configurations as rare, especially since those same interferometric surveys have succeeded in detecting the key alternative: stripped subdwarf companions over a range of luminosities (see Sections~\ref{sec:Be_sdO} and ~\ref{sec:bloated_sdO}).

Some recent works have started to discover and characterize pre-interaction systems with rapidly-rotating OB star primaries in short orbital periods. \citet{naz23} describe three massive (O7, B0, and B3) rapid rotators ($v$sin$i$ $>$ 200 km s$^{-1}$) in short period (days) binaries with pre-MS companions, indicating that these are young pre-interaction systems where the rapid rotation of the primary is an intrinsic property of the star. Other qualitatively similar examples include the rapidly rotating ($v$sin$i$ = 216 km s$^{-1}$) early-B primary in HD 149834 with a $\sim$1.2 M$_{\odot}$ companion \citep{sta21}, HD 165246 with an O8 primary with $v$sin$i$ = 243 km s$^{-1}$ \citep{joh21}, and $\gamma$ Lupi A (B5) with $v$sin$i$ = 236 km s$^{-1}$ \citep{jer21}. Others include HD 163892 \citep[O9.5IV(n), ][]{mah22} and HD 152200 \citep[O9.7IV(n), ][]{bri24}, all with $v$sin$i$ $>$ 200 km s$^{-1}$.
A dozen more B+pre-MS cases have been reported in \citet{2025arXiv251015393N} and the vast majority of them also show supersynchronous rotation.
While none of these systems include Be stars, their existence demonstrates that binary interaction is not required for achieving rapid rotation.
Some other works indeed argue that a significant fraction of B-type stars emerge on the ZAMS with a relatively high rotational velocity \citep[e.g.][]{2010ApJ...722..605H, 2022NatAs...6..480W}, but the true initial rotational velocity distribution for massive stars remains difficult to ascertain, as it requires distinguishing between stars that have evolved as single versus those that have experienced binary interaction \citep{2013ApJ...764..166D}.

\subsection{Interacting systems}

There are many types of interacting binaries across the HRD, but the ones that concern us here are primarily the massive Algol systems.  Algol-type binaries are the class of eclipsing binary stars comprising an early-type MS star and a late-type, secondary sub-giant or giant star that fills its Roche lobe. The evolutionary state of Algol systems was first explained by a mass transfer model more than 70 years ago \citep{cra55,pla70}, but details concerning the mass transfer process have slowly emerged only over the past 40 years \citep[][and references therein]{pet22}.  The \textit{massive} systems typically have total masses greater than $5~ M_\odot$.  For the B-type systems, the Roche lobe-filling (secondary) star, usually has a mass less than $2~M_\odot$. Secondaries in the O-type systems often show masses greater than $10~M_\odot$.  Less than one-third of the semi-detached binaries listed in recent catalogs \citep[e.g.][]{mal20} are massive Algols according to our definition.
X-ray emission is typically observed in Algol systems and is due to activity on the surface of the late-type giant \citep[e.g.,][]{chu04}. Highly ionized species (e.g. {\sc{N V}, \sc{C IV}, \sc{Si IV}}) tend to be seen in absorption or emission at the phase interval 0.4-0.6 during the mid-latter stages of post-mass reversal mass transfer and are likely a result of shock heating from the impacting gas stream \citep{petpol84}.

It is reasonable to suspect that the massive Algols may eventually evolve into OBe/n + sdO systems. For example, the Algol binary RY Per (B3Ve + F7II-III), which presently has a systemic mass of $8 M_\odot$  \citep{bar04}, could become a system like 28 Cyg \citep{kle24} or 59 Cyg \citep{pet13}, but its orbital separation would have to increase by a factor of 13.5 or 4.1 to resemble 28 Cyg or 59 Cyg, respectively.  The RY Per system is approaching mass transfer shutoff and the donor star is already less than $2~ M_\odot$, so there is not that much mass left to transfer. Furthermore, the rotational velocity of the mass gainer is only 215 km~s$^{-1}$, which is less than half the critical velocity needed to form the Be star’s disk but still significantly faster than if the star were tidally synchronized. The Algols may be on their way to becoming Be/n + sdOB binaries but the angular momentum transfer needs further study.

A (limited) number of systems with an accreting B star paired to a red giant have also been spotted. They are interpreted as binaries in which mass transfer is currently occurring, i.e. in which the emission line arises from active accretion of material from the companion by the B star. \citet{har15} lists nine such systems, with periods in the range 6 to 36 d and low mass ratios of $\sim$ 0.06 to 0.30 (as expected after mass ratio reversal although UX Mon, with a ratio of 1.15, is a rare pre-reversal case). There are also other examples such as, AX Mon (Be+KIII, $P\sim232$d, \citealt{eli97}) and V644 Mon (B1 V+K0 I-II, $P\sim121$d, \citealt{auf94}). Table~8 in \citet{har15} lists ten other systems that he describes as possible semi-detached emission-line binaries. The B-type mass gainers in these systems rotate significantly faster than the orbital period, but in general seem sub-critical.

\subsection{Post-interaction systems} \label{sec:postinteraction}

In post-interaction systems that now contain a Be/n star, it is generally understood that the originally more massive star transferred mass and angular momentum causing the spin-up of the mass gainer eventually leading to the rapidly rotating B star becoming a Be/n star and leaving a stripped-envelope companion star (the mass donor). There are several classes of post-interaction systems that are observed. These are described in the following subsections, which are listed roughly in order of increasing luminosity of the stripped star, followed by post-supernova systems.

\subsubsection{Supersoft X-ray emitters} \label{sec:softBeXRBs}

In recent years, a dozen OBe stars (mostly belonging to the Magellanic Clouds) were found to be associated with bright, flaring, and very soft X-ray sources \citep[e.g.][]{2018ApJ...862..167C}. Since the X-ray emission properties appear typical of nuclear burning on a WD surface, these systems were identified as Be+WD binaries. Orbital periods could be determined only in two cases: 17d for both Swift J011511.0-725611 \citep{ken21} and CXOU J005245.0-722844 \citep{gau24, mar25}. The optical components of these two systems are early-type Be stars. 
Although apparently rare, there are also binary systems with fast-rotating B-type stars (plausibly Be) that present nova behavior, where the accreting star is necessarily a WD \citep{2025ApJ...991..174C, 2025arXiv251116594S}.

\subsubsection{Low-luminosity low-mass companions} \label{sec:Be_sdO}

The post-interaction system Regulus includes a near critically-rotating (but diskless) MS B8 star, and a low-luminosity pre-WD on a circular 40 d orbit. Although the binary nature was known for some time \citep{gie08}, detecting the stripped companion required a considerable quantity of high-SNR spectroscopy due to the very low luminosity of the $\sim$0.3 M$_{\odot}$ stripped star \citep{gie20}. 
A similar system is KOI-81, found to be a rapidly rotating diskless B8V star in a circular 23.88 d eclipsing binary with a $\sim$0.2 M$_{\odot}$ stripped star \citep{mat15}. 
These are the only two near critically-rotating Bn stars with published stellar and binary parameters that we are aware of.

The relatively well studied class of Be+sdOB binaries now includes about 20 members \citep[e.g.][]{gie98, pet13, pet16, wan18, wan21, wan23}. The stripped stars are generally understood to be helium-fusing hot subdwarfs with temperatures similar to main sequence O-type stars ($\gtrsim$40 kK) and masses of about 0.5 -- 2 M$_{\odot}$. While hot, their small size ($\sim$0.5 R$_{\odot}$) makes them difficult to detect and characterize against the glare of their much more luminous MS companions.
Be+sdO binaries tend to have near-circular orbits, but in 60 Cyg and 59 Cyg the eccentricities are 0.2 and 0.14, respectively \citep{pet13, kle24}. Orbital periods range from about 30 -- 300 d. In all but two such systems ($\kappa$ Dra and HD~92406), the rejuvenated component is an early-type Be star. $\kappa$ Dra (B6) has a slightly cooler-than-typical subdwarf with a temperature typical of a MS B star, making this the first Be+sdB type system discovered \citep{kle22}. HD~92406 is an eclipsing system first reported by \citet{2018BAVSR..67...41H} with a late-type Be star and a $\sim$0.5 M$_{\odot}$ stripped star that lies between the temperature and luminosity of the stripped star in $\kappa$ Dra and those from the bulk of the known Be+sdO population \citep{2025arXiv250720053D}.

\subsubsection{High-luminosity low-mass companions} \label{sec:bloated_sdO}

Stripped stars of around one solar mass can have a similar temperature and size (and therefore luminosity) as the much more massive mass gainer during a short-lived evolutionary phase after the cessation of mass transfer \citep[e.g.][]{vil2023,riv24}. These ``bloated'' stripped stars tend to have slow rotation and near-circular orbits, although it is possible, for reasons yet unclear, that high eccentricities are allowed \citep[e.g. HIP 15429,][]{mul25}. It is presumed that in such systems, the stripped star will contract to become a core helium burning subdwarf, joining the more populated class of Be+sdOB binaries (Section~\ref{sec:Be_sdO}).

\subsubsection{High-luminosity higher-mass companions} \label{sec:highL_stripped}

At higher stripped star masses ($\sim$2 -- 8 M$_{\odot}$), the stripped star can be more luminous than the mass gainer, appearing as a lower-mass analog of Wolf Rayet (WR) stars. Such objects are expected to remain highly luminous until exploding as a supernova, rather than evolving to become a low-luminosity hot subdwarf \citep{hov25}. 
About a dozen such systems have been detected in the Magellanic Clouds \citep{got17,dro23,ram24}. One qualitatively similar Galactic case is known, HD 96670, with a stripped star of $\sim$4.5M$_{\odot}$ and an  evolved O-type star in an orbital period of only 5.28 d \citep{naz25}. The mass gainer in HD 96670, with $v$sin$i\sim$175 km s$^{-1}$, may not be a `proper' rapid rotator, but may have been spun down by some combination of evolution, AM loss through winds, and tidal effects. The temperature, luminosity, and mass of the stripped star in HD 96670 are similar to the sample from the Magellanic Clouds, although the O-type mass-gainer dominates the total flux of the system. The full sample of \citet{dro23} includes stripped stars with presumed main-sequence companions with undetected, moderate, and large contributions to the optical flux. It should be noted that the stripped stars analyzed in \citet{got17}, which is the sub-sample of systems from \citet{dro23} that are without detected companions, possibly underwent a common envelope phase where the presumed main-sequence lower-mass companion did not accrete much material or angular momentum. More detailed information about the systems with stripped stars and signatures of a luminous MS star (which may be post mass-transfer binaries) from \citet{dro23} are not currently available.

\subsubsection{Wolf-Rayet companions} \label{sec:WRs}

Classical Wolf-Rayet (WR) stars are  hot envelope-stripped stars with high initial masses in a late stage of evolution. Some fraction of WR stars have lost their envelope via binary interaction, while others (especially with relatively high initial masses) may strip their envelopes via strong winds and/or eruptive mass loss episodes \citep{2024arXiv241004436S}. The relative frequency of these two evolutionary paths for producing WR stars is still an open question \citep{2014ApJ...789...10N, 2019A&A...627A.151S, 2020A&A...634A..79S}.

For WR stars that have been stripped via Roche lobe overflow (RLOF), the mass gaining component, which is typically an O-type star, may be spun up in the process. Typical rotation rates of these mass gainers are super-synchronous, where typical post-RLOF orbital periods may be anywhere from days to months, with rotation rates found to be about $\sim$50\% of critical \citep[e.g.][]{2017MNRAS.464.2066S, 2020MNRAS.492.4430S}.

However, \citet{2018MNRAS.478.3133R} argue that the rotation rates of the O stars in WR+O binaries can be systematically overestimated by a factor of $\sim$2, due to the complexity of the composite spectra of such systems and difficulties in identifying the pseudo-continuum from which to measure absorption line properties. Nevertheless, evidence suggests that the O-type companions of binary-stripped WR stars rotate faster than apparently single O-type stars, but perhaps not much more than $\sim$30\% of critical \citep[barring some exceptions; ][]{2019A&A...627A.151S}. Some physical processes, such as magnetic braking during the mass transfer phase \citep{2018MNRAS.478.3133R} and/or tidal interactions, act to quickly slow the rotation of the mass- and AM-gaining O-type star.

\subsubsection{Be/X-ray binaries} \label{sec:BeXRBs}

Be/X-ray binaries (BeXRBs) are a subclass of Be binaries made up of a Be star and a neutron star \citep[NS;][]{rei11}. The primary stars in these binaries are always early-type Be stars \citep{rei11}.  As emitters of high energy X-rays, they represent the largest class of high-mass X-ray binaries \citep{for23}. They are among the brightest X-ray sources in the sky \citep{rei11}, with X-ray luminosities on the order of 10$^{36-37}$ erg s$^{-1}$ or larger. The X-ray flux is produced by the accretion of disk material onto the NS. Both the X-ray and visible flux in such systems are highly variable, and orbits are often eccentric and misaligned with respect to the Be star’s equator \citep{hug99, mar11, sal20, mar24b}. It is generally accepted that the supernova event that creates the NS in such a system produces a kick that also sets the NS on a new, disrupted orbital path. \citet{mar09a} found that while such a kick can explain the observed misalignment angles of BeXRBs, the eccentricities may be influenced by more complex events (such as circularization over the lifetime of the system). There are more than 70 known BeXRBs in the Galaxy \citep{for23}, with roughly 100 more found in the SMC and LMC \citep{coe15, ant16}. It is possible that Bn stars may exist in similar binary configurations, however they will not emit in X-rays due to the lack of circumstellar material (the accretion of which powers the X-ray emission in BeXRBs).

Black Hole (BH) companions, which may have very low X-ray luminosities \citep{2024A&A...690A.256S}, are also starting to be discovered orbiting massive OB stars, which at least in some cases rotate rapidly \citep{2022NatAs...6.1085S}. The system MWC 656 was claimed to be a Be+BH binary \citep{2014Natur.505..378C}, but this has been challenged by more recent analysis of this system which instead suggests the companion to the Be star is consistent with being a hot stripped star \citep{2023A&A...677L...9J, 2024IAUS..361..332R}. To our knowledge there are currently no confirmed Be+BH binaries in the literature.

\subsubsection{Runaway Be stars} \label{sec:runaways}

Runaway stars (or `walkaway stars', being less extreme but qualitatively the same) are systems with anomalously high space velocities and/or found at large distances from their birth environment compared to what is expected from typical kinematics. Generally, runaways were imparted with high space velocities either through dynamical scattering (e.g. multiple-body interactions in a dense cluster environment), or by having a binary companion that explodes as a SN. Some fraction of Be stars are runaways, although the statistics depend on the environment being studied. \citet{bou18}, for instance, determined that about 13\% of Galactic Be stars are runaways, and interpreted this as evidence for a high fraction of Be stars being post-interaction products. \citet{dor20} analyzed the runaway populations of the Small Magellanic Cloud and reached a similar conclusion -- a high fraction of runaways with high $v$sin$i$ (including Be stars) suggests that binary interaction is a major contributor to rapid rotation. However, it should be emphasized that rapidly-rotating runaways are not all guaranteed to have been spun up by binary interaction, since dynamical scattering can be a major contributing factor to the runaway population in general \citep{car23}.

\section{Comparing binary populations} \label{sec:compare_pops}

It is now well established that some fraction of Be/n stars have been spun up via binary interaction as demonstrated by the growing number of confirmed Be+sdO and similar post-interaction systems (Section~\ref{sec:postinteraction}). 
We have attempted to compile information from the literature for as many known post-interaction Be/n binaries as possible in Appendix~\ref{secA1}. However, these lists are probably not exhaustive, and this is a rapidly advancing field such that there are likely to be newly discovered systems and improved/updated parameters by the time of publication. 

The large majority of confirmed post-interaction systems with a Be star (including all BeXRBs and runaways) contain an early-type Be star ($\sim$B3 and earlier), and a stripped star mass of about 0.5 M$_{\odot}$ and above.
One aim of this section is to test whether or not there is reason to believe there is a similar population of post-interaction late-type Be/n+stripped binaries which have so far avoided direct detection. 
Another is to compare the SB1 Be/n population to the confirmed Be/n+stripped population.
In this Section, only systems that could plausibly be post-interaction are considered. That is, very wide binaries (visual doubles, or with orbital periods on the order of many years) are ignored.

\begin{figure*}[h]
\centering
\includegraphics[width=0.75\hsize]{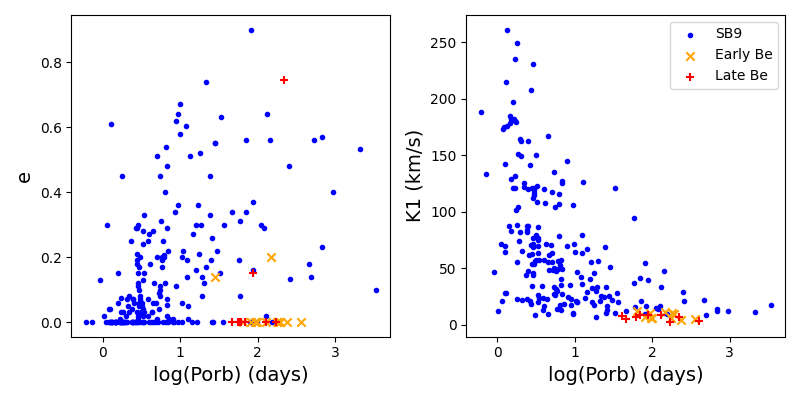}
\caption{Eccentricity (left) and radial velocity (right) vs. orbital period for main sequence B-type stars from the SB9 catalog. Early-type and late-type Be/n stars are indicated as in the legend. The two early-type outliers in the left panel are 59 Cyg and 60 Cyg, and the two late-type outliers are Pleione and 88 Her (see Appendix~\ref{secA1}). Of these, 88 Her is the only case where the eccentricity may be unreliable. }
\label{fig:SB9_comparison}
\end{figure*}

First, it is useful to compare certain binary parameters measured for Be/n stars to those of the normal B star binary population. The most readily available orbital parameters are the period (P$_{\rm orb}$), eccentricity ($e$), and radial velocity semi-amplitude ($K_{\rm 1}$), as these can be measured for SB2 and SB1 systems. Figure~\ref{fig:SB9_comparison} plots $e$ and $K_{\rm 1}$ versus P$_{\rm orb}$ for main sequence (non-Be/n) B stars from the SB9 binary catalog \citep{pou04}, and for the early and late Be/n binaries for which this information is currently available (see Appendix~\ref{secA1}). Although there are several additional factors that contribute to the SB9 sample (a range of secondary masses, selection biases, etc.), it is evident that the Be/n binary population stands apart. Be/n binary orbital periods are found only between about 30 -- 300 d\footnote{Excluding the two OBe+WD binaries with shorter orbits, Sec.~\ref{sec:softBeXRBs}, and Sk -71 35 which is likely still undergoing mass transfer, Tbl.~\ref{tbl:tbl_highM}.}, even though shorter periods are in principle easier to detect (while the absence of longer periods may be partly a detection bias). With a few exceptions, the eccentricity of Be/n orbits is very close to zero, which is rare for main sequence B star binaries in the period range occupied by the Be/n stars. The RV semi-amplitudes for Be/n stars are consistently low, implying lower-mass companions compared to the SB9 sample over this period range. In this view, there are no major differences between the early- and late-type Be/n binaries, but overall the Be/n binaries differ from the B stars in general.

\begin{figure*}[ht]
\centering
\includegraphics[width=0.9\hsize]{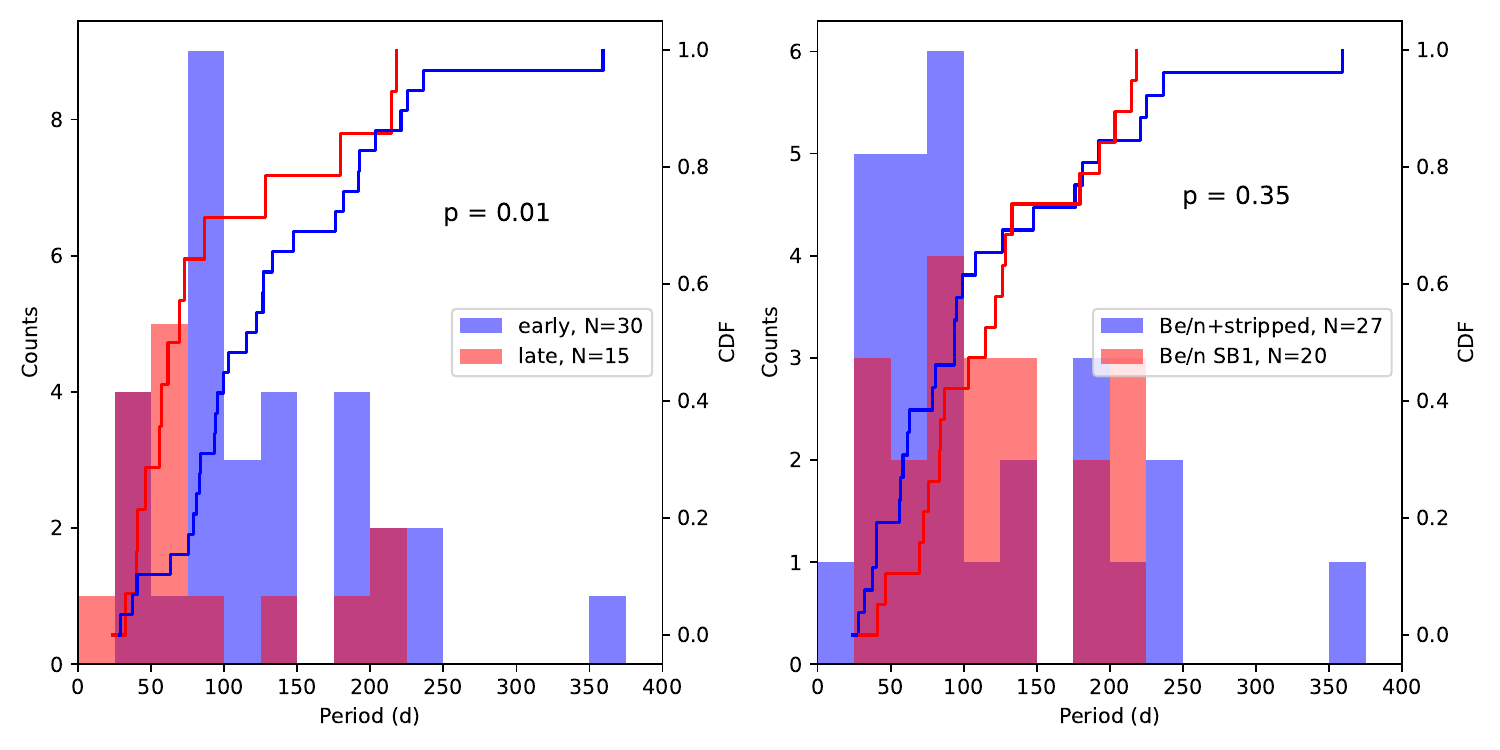}
\caption{\textit{Left:} Histogram of orbital periods for early- (blue) and late-type (red) binaries with Be/n star primaries, including both SB1 systems and those containing confirmed stripped stars. \textit{Right:} Histogram for Be/n binaries confirmed as post mass-transfer (with directly detected stripped stars, blue) and for SB1 Be/n binaries (red).  
The cumulative distribution function (CDF) is plotted as solid lines for these two samples. The $p$ value of the two-sided Kolmogorov-Smirnov comparing each pair of samples is given in both panels.}
\label{fig:compare_Porb}
\end{figure*}

Figure~\ref{fig:compare_Porb} compares the distribution of orbital periods for the known early- and late-type Be/n binaries (left panel). 
The $p$ value of the two-sided Kolmogorov-Smirnov (KS) test is 0.01, indicating that if the two samples are drawn from the same distribution the probability of obtaining a KS statistic at least this extreme is 1\%.
However, given the inhomogeneous data and unknown sampling biases, no inferences can be made from this simplistic comparison.
Although the range in periods is similar, at present there is a relative excess of late-type Be/n binaries at shorter orbital periods compared to early-type Be/n binaries. Or, perhaps, there is a dearth of detected longer-period late-type Be/n binaries as a consequence of selection effects, especially since less observational resources have historically been applied to late-type Be/n stars.

The right panel of Figure~\ref{fig:compare_Porb} compares the orbital periods for confirmed post-interaction Be/n+stripped star binaries (Tbls.~\ref{tbl:tbl_sdOB},~\ref{tbl:tbl_bloated},~\ref{tbl:tbl_highM}), with the orbital periods of Be/n binaries where the nature of the companion is not yet known (the SB1 sample, Tbl.~\ref{tbl:tbl_highM}). In this comparison, the distribution of orbital periods of these two samples are consistent with being identical (the $p$ value of the two-sided Kolmogorov-Smirnov test is 0.35).

\begin{figure*}[h]
\centering
\includegraphics[width=0.45\hsize]{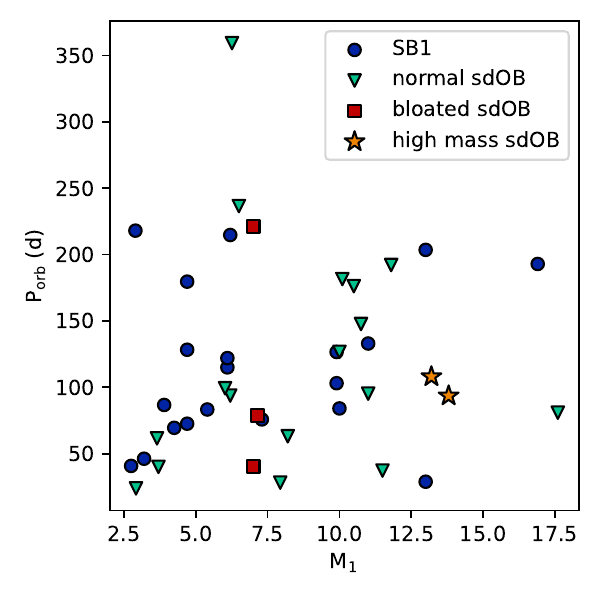}
\includegraphics[width=0.45\hsize]{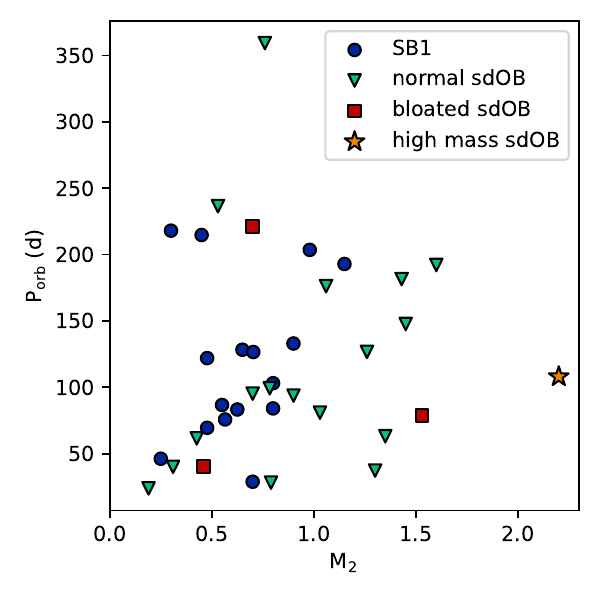}
\includegraphics[width=0.45\hsize]{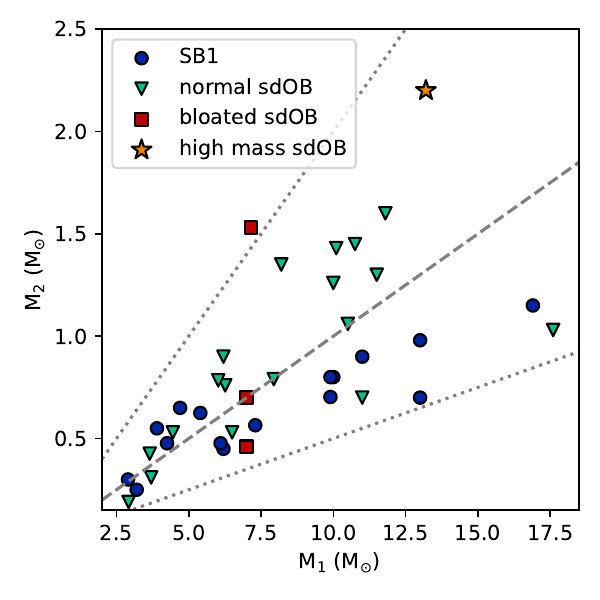}
\caption{\textit{Top-left:} Comparison of the mass of the primary Be/n stars (M1) to the orbital period. \textit{Top-right:} Comparison of the mass of the lower-mass secondary star (M2) to the orbital period. \textit{Bottom:} Comparison of the secondary to primary mass. In the bottom panel, the dotted lines correspond to mass ratios of 0.2 and 0.05, and the dashed line to 0.1. In all panels, systems containing secondaries whose nature is unknown (SB1), and secondaries confirmed as being normal stripped stars, bloated stripped stars, and high mass stripped stars are indicated with different symbols. We further note that B4 and later types correspond to stars of $\sim$ 5 M$_{\odot}$ and lower. Two systems, 2dFS 163 ($q = 0.4$) and 2dFS 2553 ($q = 0.5$), both containing a high-mass stripped star and an early-type Be/n star \citep{ram24} are not shown in these figures.}
\label{fig:Mstripped_vs_Porb}
\end{figure*}

Figure~\ref{fig:Mstripped_vs_Porb} investigates relationships between primary and secondary masses and orbital periods. 
The top two panels in Figure~\ref{fig:Mstripped_vs_Porb} examine correlations between the orbital periods and the primary (left) and secondary (right) masses. There is significant scatter in both correlations, 
and it is not clear that there is any correlation between orbital period and primary and secondary masses.
The bottom panel of Figure~\ref{fig:Mstripped_vs_Porb} shows a strong correlation between the mass of the secondary and that of the Be/n star. 
There does not seem to be any qualitatively different behavior in systems with early-type Be/n stars (with primary masses $\gtrsim$6 M$_{\odot}$) compared to late-type.
Neither are there any glaring differences in the above correlations between systems of different types (e.g. SB1 systems and those with more- or less-luminous stripped stars). However, in the comparison between primary and secondary masses, the SB1 systems including early-type Be/n stars tend to have somewhat lower estimated secondary masses, which again may be a selection effect as lower-mass stripped stars are generally of lower luminosity and thus harder to detect.

\begin{figure*}[ht]
\centering
\includegraphics[width=0.99\hsize]{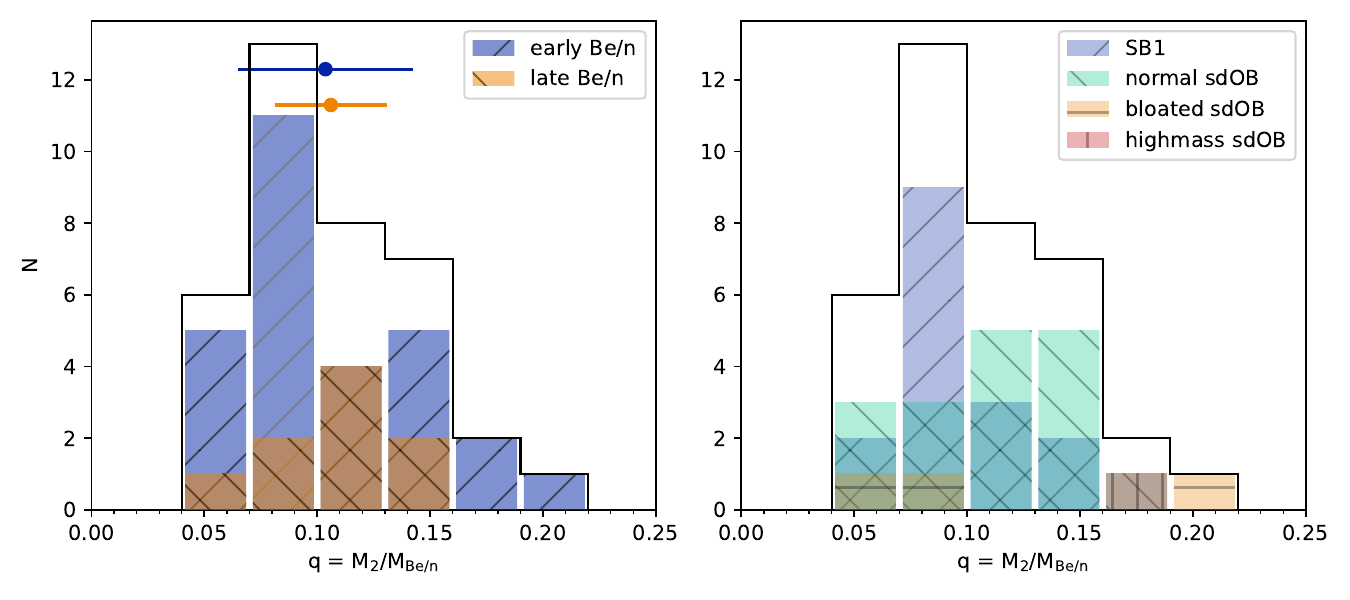}
\caption{ \textit{Left:} Histogram of mass ratios (M$_{\rm stripped}$/M$_{\rm Be}$) for the early- and late-type binaries, and their sum (black solid line). The mean and standard deviation for the early- and late-type mass ratios are plotted, with values of q = 0.10 $\pm$0.04 (early), and q = 0.10 $\pm$0.03 (late). \textit{Right:} The same, but divided into the four different categories of binary systems as indicated in the legend, but without regard to spectral type. }
\label{fig:hist_of_q}
\end{figure*}

Figure~\ref{fig:hist_of_q} (left panel) plots a histogram of the mass ratio ($q$) for the same sample as in Figure~\ref{fig:Mstripped_vs_Porb} showing that the early- and late-type populations are similar. The mean values and their standard deviations are $q = 0.10\pm0.04$ (early) and $q = 0.10 \pm0.03$ (late). The right panel of Fig.~\ref{fig:hist_of_q} is similar, but comparing the different categories of binaries. There are two outliers with less extreme mass ratios (omitted from the calculation of the mean $q$ values and not shown in Fig.~\ref{fig:hist_of_q}). These are the systems 2dFS 163 ($q = 0.4$) and 2dFS 2553 ($q = 0.5$), both containing a high-mass stripped star and an early-type Be/n star \citep{ram24}.

From the above, a few inferences can be drawn. There appear to be no major differences between Be/n binaries when comparing early versus late spectral types in terms of orbital period, eccentricities, mass ratios, and trends between these quantities. 
However, the orbital period distribution for late-type Be/n binaries may skew towards shorter periods, although this may be related to a selection bias.
The same can be said when comparing confirmed post-interaction Be/n binaries with stripped stars, and the SB1 systems where the nature of the secondary is unknown. 
It therefore is reasonable to assume that a significant fraction of late-type Be binaries are post-mass transfer products (as well as early-type SB1 systems), with the secondary being a stripped low-mass star, rather than a low-mass MS star. 
From an observational point of view, the problem remains that very few stripped stars have yet been directly detected orbiting late-type rapid rotators, and there are several very well studied early-type binaries, especially the $\gamma$ Cas analogs \citep[early-type Be stars with characteristic X-ray emission,][]{naz18, rau22}, whose companions still evade direct detection. Potential reasons for this are explored in the next section.

\section{Be/n binaries on the Hertzsprung–Russell diagram} \label{sec:BPASS}

\subsection{Observations}

\begin{figure*}[ht]
\centering
\includegraphics[width=0.95\linewidth]{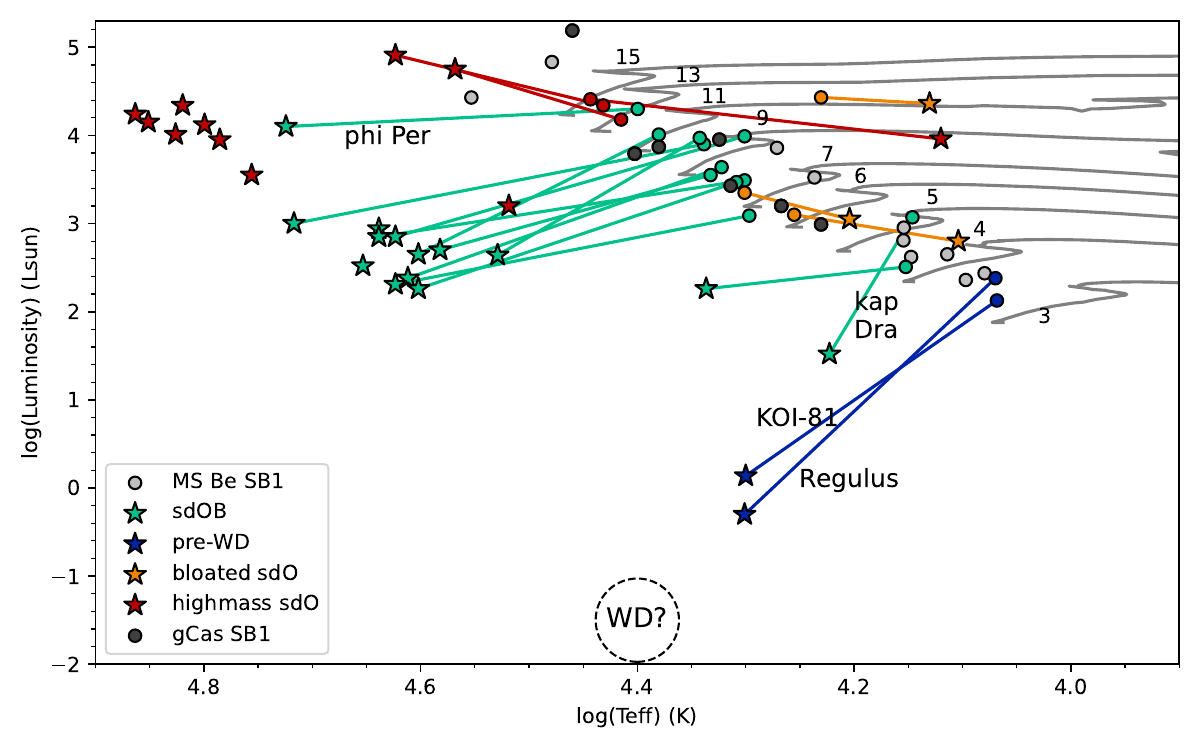}
\caption{Observational Hertzsprung–Russell diagram for known Be/n binaries. The filled circles represent the near-MS rapidly-rotating B-type stars, and the star symbols represent stripped stars. Whenever the luminosity and temperature of both components for a Be/n+stripped binary are known, the location of the MS star and stripped star are connected by a solid line. Different colors represent different classes of stripped stars, as represented in the legend. SB1 Be/n stars are indicated by light gray circles, and the SB1 systems that are $\gamma$ Cas analogs are indicated by dark gray circles.
MS evolutionary tracks from the Geneva stellar models \citep{geo13} from 3 -- 15 M$_{\odot}$, with rotation rates at 95\% of critical (here defined as $\Omega$/$\Omega_{\rm crit}$) and Z = 0.014 are plotted as gray lines. All high-mass sdO stars are from the Magellanic clouds, with the remainder of systems being Galactic.
}
\label{fig:HRD}
\end{figure*}

While the rapidly-rotating mass gainers in Be/n+stripped binaries populate the near-MS, the stripped stars are found in several different regions in the Hertzsprung–Russell diagram (HRD). The temperature and luminosity of a stripped star depends primarily on its mass and evolutionary stage (i.e. how much time has passed since the end of mass transfer). 
Figure~\ref{fig:HRD} examines the temperature and luminosity of near-MS Be/n stars (primarily Be) and their stripped companions in the HRD, whenever such measurements or estimates are available. 

The most well-observed subset of non-NS stripped stars are the helium-burning sdOB stars (Sec.~\ref{sec:Be_sdO}). These are generally found along the `helium main sequence', running roughly parallel to the MS, but offset at lower luminosity and higher temperature. This helium main sequence is probably a continuum that spans from the lowest-mass to highest-mass stripped stars, although studies with differing science objectives often focus on specific regions.

The first stripped star that was directly detected orbiting a Be star is in the system $\phi$ Per \citep{gie98}. It is the hottest and most luminous stripped star among the `intermediate mass' sdOB population. Many other qualitatively similar systems (but with less luminous sdO components) have since been discovered (Sec.~\ref{sec:Be_sdO}).
The sdB star in $\kappa$ Dra has the lowest luminosity (and lowest temperature) of this population \citep{kle22}. Fig.~\ref{fig:HRD} illustrates an obvious lack of confirmed sdB stars, which should be located somewhere between the stripped star of $\kappa$ Dra and the remainder of the sdO components. 

The bloated stripped stars (Sec.~\ref{sec:bloated_sdO}) are near-indistinguishable from MS B stars in the HRD. Their very low surface gravity, as inferred through spectroscopic analysis and/or their dynamical mass is what sets them apart as low-mass stripped stars \citep{bod20b,riv24}. It is generally understood that these will quickly evolve to the He MS after their short-lived `bloated' stage. Due to their high luminosity and narrow lines (being slow rotators), they are relatively easy to discover, but intrinsically rare as this is a short evolutionary stage.

The high-mass stripped stars (Sec.~\ref{sec:highL_stripped}) can be found at similar or higher luminosity than their MS companion. Considering the broad lines of the mass gainer in these systems, the typical challenge is reversed -- it can be more difficult to obtain spectroscopic quantities of the MS mass gainer. Although several high-mass stripped stars are now known from the SMC and LMC (see Sec.~\ref{sec:highL_stripped}), these are recent discoveries and in most cases the mass gainer has not been characterized. Nevertheless, these stripped stars lie along the same helium main sequence as the sdOB population. 

Finally, the lowest-mass stripped stars ($\lesssim$0.4 M$_{\odot}$) are unable to fuse helium in their core (Sec.~\ref{sec:BPASS_predictions}). After their envelope has been stripped and there is insufficient hydrogen available for fusion, they evolve towards becoming a WD and do not loiter on the He MS. The two known examples (the lowest-luminosity stripped stars in Fig.~\ref{fig:HRD}) are $\sim$2 orders of magnitude fainter than any other known stripped stars with rapidly-rotating B-type primaries. 

One population that is perhaps missing from the observational HRD of Fig.~\ref{fig:HRD} are `proper' WDs, with luminosities even less than the two known pre-WDs. Perhaps counter-intuitively, these may evolve from relatively high-mass stripped stars in a `sweet spot' -- too low mass to explode as SNe (leaving behind a NS; $\lesssim$2.6 M$_{\odot}$), but massive enough that they evolve relatively quickly and can exhaust their supply of helium before their companion evolves off the MS (at which point it will no longer be identifiable as a rapid rotator). Although no such WDs have been observed directly, there is circumstantial evidence that suggests that at least some of the $\gamma$ Cas analogs  have WD companions \citep{gie23}. There is additionally the population of likely Be+WD systems in the Magellanic Clouds (Sec.~\ref{sec:softBeXRBs}), although these cannot be placed in the HRD. 
This remains an active and challenging field of study, as the origin of the anomalous X-ray emission of $\gamma$ Cas stars is still unknown. Not only would such WDs have a low intrinsic luminosity, but those with massive Be hosts (e.g. the $\gamma$ Cas analogs) would have such an extreme contrast ratio as to make the WDs undetectable in UV/visible spectroscopy even in principle.

\subsection{Binary model predictions} \label{sec:BPASS_predictions}

If the companions to the SB1 Be stars plotted in Figure~\ref{fig:HRD} are stripped stars, where would they be located on the HRD? To investigate this question, we used the Binary Population and Spectral Synthesis code \citep[BPASS,][]{eld17} and the Hoki package \citep{ste20} to examine evolutionary tracks for systems that would undergo mass transfer and result in (presumably rapidly rotating) B stars on the MS and lower-mass stripped stars. When filtering the BPASS models, we searched the binary models at solar metallicity (Z = 0.02), with an initial orbital period of 100 days or less, that must have undergone mass transfer resulting in a B-type star (3 to 18 $M_\odot$) with a low mass companion (0.2 to 5 $M_\odot$), the ranges of which are consistent with the observed masses in Figure~\ref{fig:Mstripped_vs_Porb}. We then also removed systems where the final orbital period was less than the initial orbital period, indicating highly non-conservative mass transfer. In total, there were 443 BPASS models which met the selection criteria.

To generate the binary population, BPASS bases the initial mass function on \citet{1993MNRAS.262..545K}, adopting a power-law slope from 0.1 to 0.5 M$_{\odot}$ of -1.3, and a slope of -2.35 at higher masses. A flat distribution is prescribed for the initial mass ratio and for the logarithm of the orbital period.

\begin{figure}
    \centering
    
    \includegraphics[width=\linewidth]{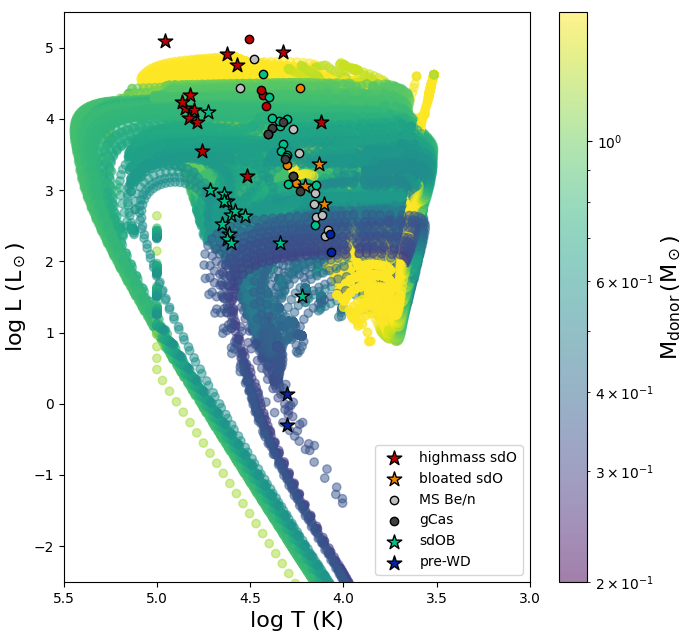}\\
    \caption{HR diagrams showing the evolutionary tracks of the donor star, coloured by mass, for the BPASS models that met our search criteria as described in Section~\ref{sec:BPASS_predictions}. The position of the observed mass gainer stars and their companions from Figure~\ref{fig:HRD} are also plotted for reference.}
    \label{fig:BPASS_HR_secMASS}
\end{figure}

\begin{figure*}
    \centering
    \includegraphics[width=0.49\linewidth]{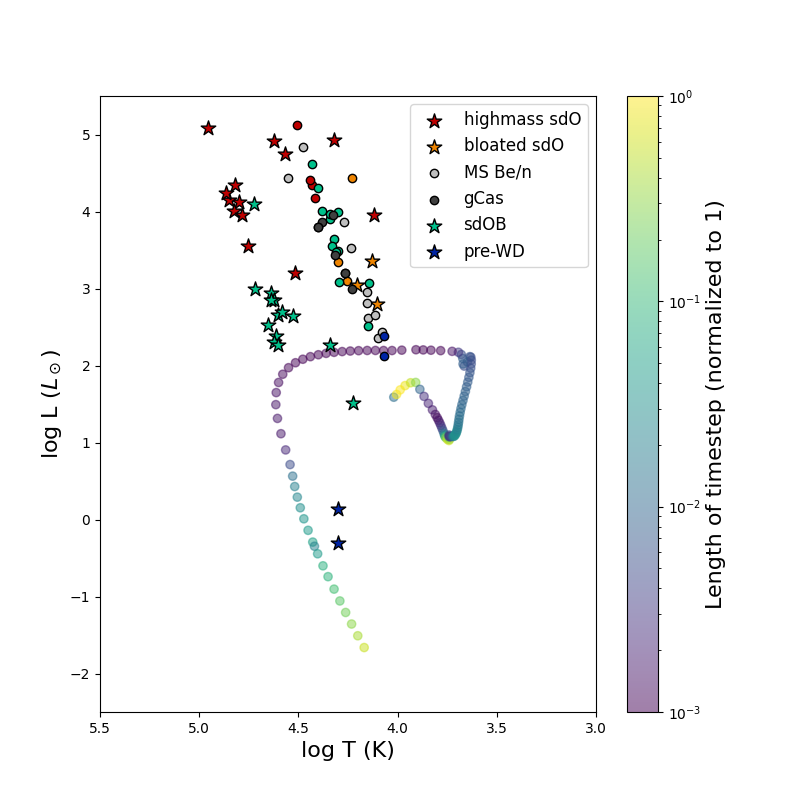}
    \includegraphics[width=0.49\linewidth]{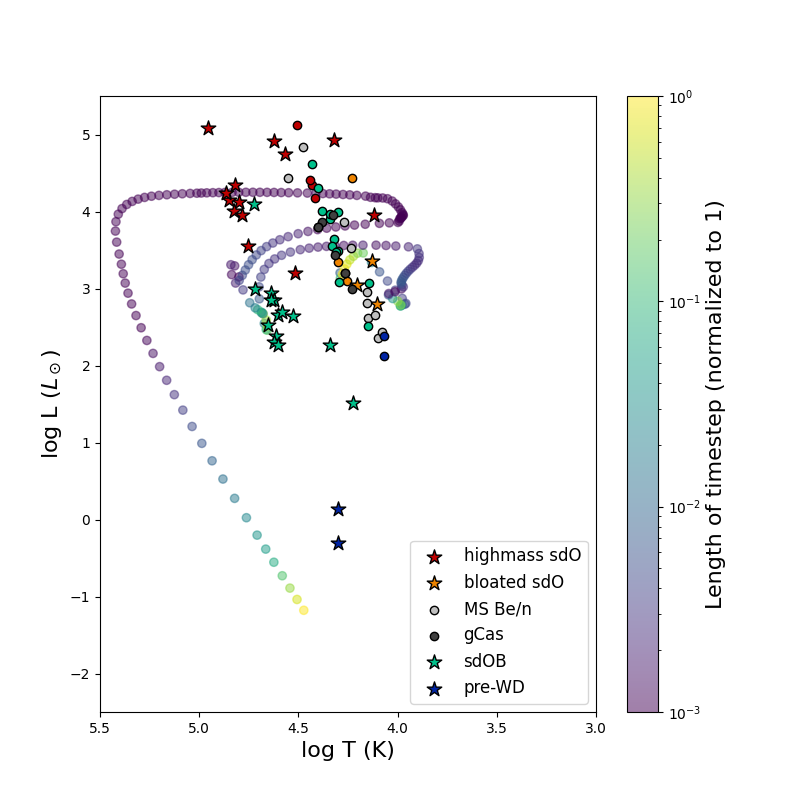}
    \caption{Example evolutionary tracks for a lower mass ($\rm 0.31\,M_\odot$) stripped star (left), and higher mass stripped star ($\rm 0.85\,M_\odot$ right), colored by normalized timestep length indicated by the color bar. The same observations as Figure \ref{fig:BPASS_HR_secMASS} are overplotted.}
    \label{fig:example_bpass_tracks}
\end{figure*}

In Figure~\ref{fig:BPASS_HR_secMASS} we plot the evolutionary tracks of the mass donors of these 443 models, colored by donor star mass, along with the observed Be/n+stripped systems. There are essentially two categories of stripped stars that are relevant. Higher mass stripped stars ($\gtrapprox$ 0.35 \rm M$_{\odot}$) undergo core helium burning and usually maintain a relatively higher luminosity for a long time on the He MS, while lower mass stripped stars ($\lessapprox$ 0.35 \rm M$_{\odot}$) cannot fuse helium and instead directly evolve along a cooling sequence. Examples of each are shown in Figure~\ref{fig:example_bpass_tracks}, where we see a $\rm 0.31\,M_\odot$ non-helium burning stripped star (left column) evolve directly onto a cooling track to become a WD, while a $\rm 0.85\,M_\odot$ stripped star (right column) fuses helium in the sdO region of the HR diagram, increases in luminosity, and then moves along a wider cooling track to eventually become a higher mass CO WD.

Figure \ref{fig:example_bpass_tracks} is colored by the normalized timestep length, for one low-mass and one high-mass donor star in the left and right columns respectively. The low-mass system started with masses of 2.5 $\rm M_\odot$ and 1.5 $\rm M_\odot$ for the mass donor and gainer respectively, and an orbital period of 1.6 days, and ended with masses 0.31 $\rm M_\odot$, 3.42 $\rm M_\odot$, and a period of 59.9 days. The high-mass system started with masses of 6.5 $\rm M_\odot$ and 3.9 $\rm M_\odot$ for the mass donor and gainer respectively, and an orbital period of 2.5 days, and ended with masses 0.85 $\rm M_\odot$, 8.16 $\rm M_\odot$, and a period of 65 days. 
The timestep-colored tracks are plotted until the approximate end of the mass gainer's MS. 
This timespan is set by the MS lifetime of the equivalent single-star BPASS model that matches the post-interaction mass of the accreting star.

We find that we can divide the post-MS evolution of the higher-mass stripped stars (i.e. those capable of sustaining core helium fusion) into three sections: the first left-moving track, between the end of the MS to the start of the helium main sequence, the right-moving track, from the start of He burning to the tip of the second low-temperature loop, and the tip of this loop to the WD stage (around $\log(L)\,=\,0$). Across our selected BPASS models that exhibit this behavior, these stages last on average 29 million years, 4.5 million years, and 1.7 million years, respectively, with the majority of this time concentrated at the start of He burning. This points to one possible reason why many companions are found in the sdO stage of their evolution -- they simply spend most of their post-MS life as a hot subdwarf prior to becoming a WD. As for the low-mass donor stars that do not go on to fuse helium, the average time from the end of MS to the WD stage (again defined here as $\log(L)\,=\,0$) is about 170 million years. Aside from the long time spent on the MS, stars on the high-mass track spend a considerable amount of time in the helium burning phase, around the region of observed sdO companions, as well as at the WD phase. The low-mass stripped star, however, does not have a long pause in this sdO region, and instead progresses to the longer WD phase (after gradually passing through the `pre-WD' phase).

\begin{figure}
    \centering
    \includegraphics[width=0.99\linewidth]{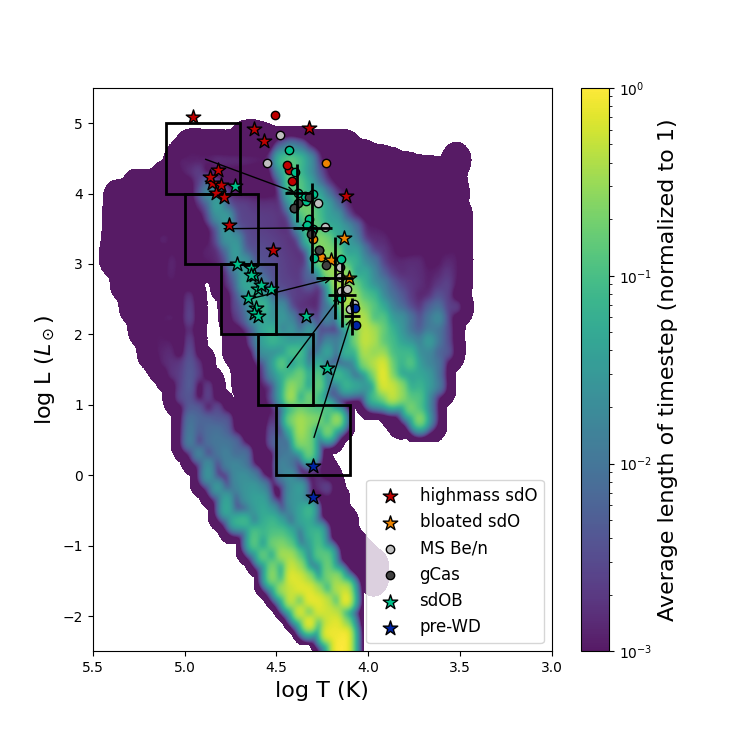}
    \caption{
    Heatmap of the 443 analyzed BPASS models, showing the average length of each timestep, normalized to 1 for each individual model. The observed Be/n and companion stars are overplotted as indicated by the legend. 
    The helium MS has been broken down into five boxes. These boxes are connected by arrows to points showing the average position of the MS mass-gainer for models that begin fusing helium in each box. The error bars indicate the standard deviation of the average mass-gainer position.
    }
    \label{fig:BPASS_age_heatmap}
\end{figure}

To get an overall representation of the relative time where stripped stars occupy different regions of the HRD in our selected models, Figure \ref{fig:BPASS_age_heatmap} shows a heatmap of where the donor stars in the 443 BPASS models are found, up to the estimated end of the gainer star MS. In this representation, we see that there are three main strips in the HRD where the donor spends the most time: the MS track (upper right), the subdwarf track or ``helium main sequence" (middle track), and the WD track (lower left).
We can see that the dichotomy of high-mass and low-mass subwarfs creates a continuous distribution between the observed high mass subdwarfs (sdO), and the few observed low mass subwarf (sdB) and pre-WD companions, stretching roughly between $4.0 < \log(T) < 4.75$ and $0.5 < \log(L) < 5$. 
The high-luminosity region of the He MS is relatively less populated, 
due to these higher-mass stripped stars being intrinsically less common and also exhausting their He fuel relatively rapidly.
We note that the number of models where the donor ends as a WD while the mass gainer is still on the MS may be overestimated due to the following two reasons. 
It is possible that the MS lifetimes of the mass gainers is overestimated, as we calculate this value assuming that the star begins on the ZAMS with its post-interaction mass. Additionally, the rejuvenated star MS lifetimes would be shorter if mass transfer is more conservative than prescribed in BPASS (as discussed later in this section).
On the other hand, this may be partly offset by the notion that rapid rotation extends the MS lifetime by enhancing mixing of H into the core, which is not accounted for in the BPASS evolution of the mass gainers.

To compare the helium-fusing stripped stars with their main-sequence companions, we have separated the helium main sequence into five sections, as denoted by the boxes in Fig.~\ref{fig:BPASS_age_heatmap}.
Within a given box, we select all donor tracks that began to fuse helium, and find the mass of the gainer star at the time of He ignition. This mass is then converted to a position on the HR diagram using the BPASS single-star models. The average position and standard deviation is plotted, with an arrow connecting each box to its average gainer position. From this method, we can see that along the helium main sequence, higher mass stripped stars will tend to have higher mass MS companions, while lower mass stripped stars will have lower mass MS companions. 

It is noteworthy that the middle box, containing the `normal' observed sdO population, points approximately to the middle of the B-star MS, while from observations we know their Be companions are of higher temperature and luminosity (Fig.~\ref{fig:HRD}). This may suggest that the observed Be+stripped population demands more conservative mass transfer than is prescribed in BPASS for this parameter space. This notion seems supported by other recent work, \citep[e.g.][]{lec25}, which indicates that the mass transfer that produces early-type Be stars is more conservative than is often assumed. 
The prescription for mass transfer rates implemented in BPASS is limited by the thermal timescale of the accretor, $ \dot{M_{2}} \leq M_{2}/\tau_{KH}$, where the mass loss rate from the primary via Roche lobe overflow is described in Eqs.~1 -- 3 in \citet{eld17}. 

\begin{figure*}
    \centering
    \includegraphics[width=0.49\linewidth]{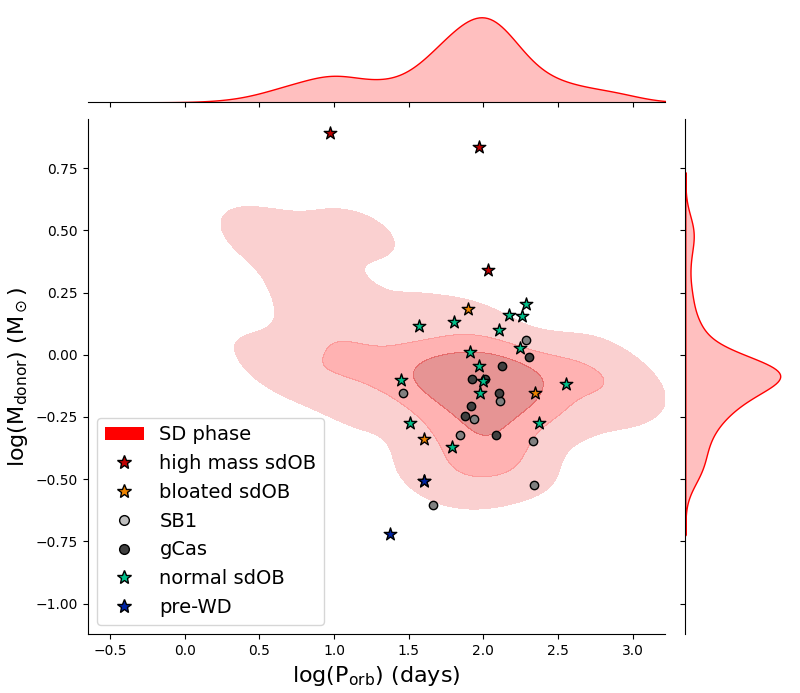}
    \includegraphics[width=0.47\linewidth]{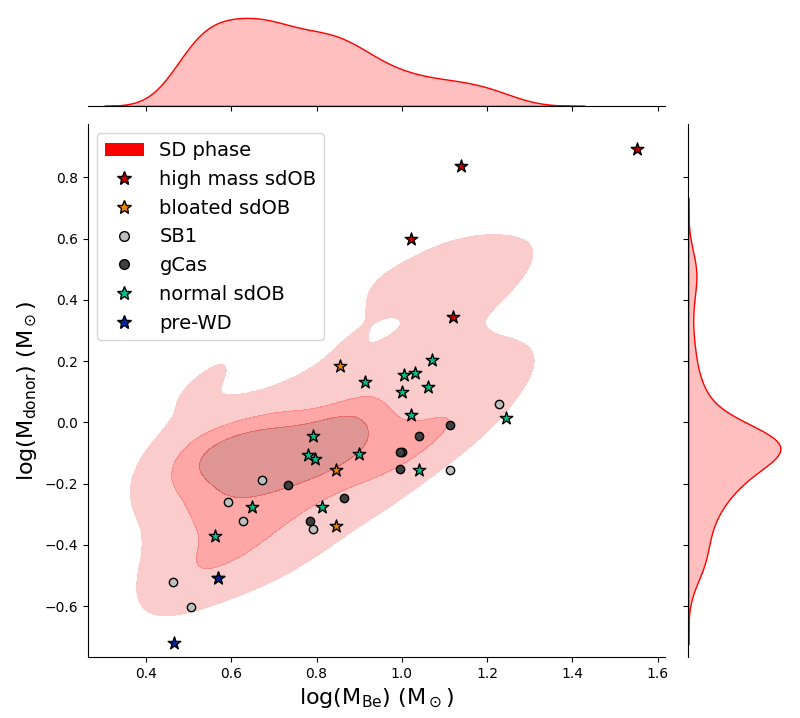}
    \caption{Left: log-log plot of orbital period versus donor star mass. Right: log-log plot of Be star mass versus donor star mass (right). Both plots show the distribution of our selected BPASS models at the subdwarf/pre-WD phase (after crossing the ZAMS from right to left after mass transfer). The marginal graphs show the 1-dimensional distribution of the $x$ and $y$ axes. Plots were made using the seaborn package \citep{seaborn}. The observed systems from Figure \ref{fig:Mstripped_vs_Porb} are also plotted on both panels.}
    \label{fig:BPASS_MvP_MvM}
\end{figure*}

Figure~\ref{fig:BPASS_MvP_MvM} shows the distribution of orbital period versus donor mass, and the gainer star (presumed to be a Be star) mass versus donor mass, at the subdwarf/pre-WD stage (after crossing the ZAMS from right to left after mass transfer), along with the observed data from Figure \ref{fig:Mstripped_vs_Porb}. 
There is no obvious correlation between the mass of the stripped star and the orbital period, neither in the BPASS models nor in the observed sample.
However, there is a clear relationship between gainer mass and donor mass in both the models and observations. 
The BPASS models also predict a short-period tail that is not observed in the Be/n binaries, as well as higher values of $q$ compared to the observed systems.

\begin{figure}
    \centering
    \includegraphics[width=0.8\linewidth]{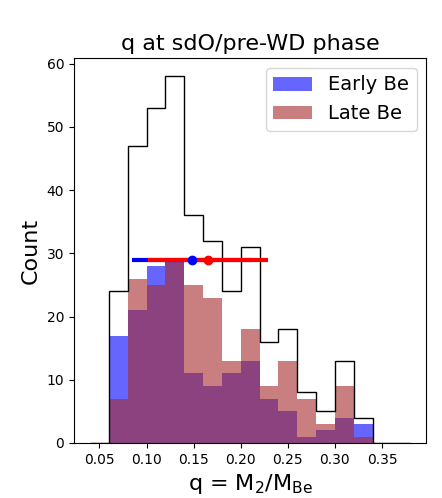}
    \caption{Histograms of the mass ratio for our selected BPASS models when the donor is at the sdO/pre-WD phase. The histograms are separated into early ($\geq$ 6 \rm $M_\odot$) and late ($<$ 6 \rm $M_\odot$) mass gainer stars, and the combined histogram is plotted in black. The mean and standard deviation of the early and late-type distributions are plotted as error bars.}
    \label{fig:BPASS_Q_hist}
\end{figure}

A histogram of the mass ratio ($q$) is shown in Fig.~\ref{fig:BPASS_Q_hist}. For each pair of stripped+B star, $q$ was calculated at the first timestep where the donor track crosses the ZAMS moving left across the HR diagram (i.e. beginning to emerge as an sdOB or pre-WD). The average values and standard deviations of $q$ for the early- and late-type B stars are $0.148\pm0.064$ and $0.166\pm0.064$, respectively.
The distribution in $q$ generated from BPASS skews towards higher values than the observed population (Fig.~\ref{fig:hist_of_q}), again suggesting that mass transfer is more conservative than prescribed in BPASS.

We have shown observationally the correlation between the mass of the stripped star and the mass gainer (Figure~\ref{fig:Mstripped_vs_Porb}). The higher mass Be stars therefore tend to have higher mass companions that spend a significant amount of the Be star MS lifetime in the core helium burning stage and appear as sdO stars. On the other hand, the companions of lower mass Be stars cool to become sdB or WD stars. 
The BPASS models predict that the stripped stars in our prescribed parameter space spend a long time fusing helium along a region in the HRD that is below but parallel to the MS (Fig.~\ref{fig:BPASS_age_heatmap}). The known stripped sdO stars lie along the upper half of this region in good agreement with the models (excluding the bloated stripped stars which are presumed to be evolving towards this region). BPASS predicts that the lower half of this region is well populated by stripped sdB stars and pre-WDs with mid- and late-type Be/n companions. However, to date there are only four such lower-luminosity stripped stars with rapidly rotating B-type companions known -- two sdB stars (with mid/late Be companions), and two pre-WDs (with late Bn companions). Observational efforts are therefore required to fill in this parameter space to paint a more complete picture of the continuum of post-interaction systems.

\section{Observational methods for detecting low-mass stripped companions} \label{sec:outlook}

The detection of stripped companions to Be stars can be done through several techniques. For NS and potential WD companions, this is done in the X-rays (Sec.~\ref{sec:BeXRBs}~\ref{sec:softBeXRBs}). UV spectroscopy has historically been the tool of choice for sdOB stars \citep{wan21, wan23}, but some systems are also detectable in optical spectroscopy \citep{cho18}. 
Bloated stripped stars are easily detectable in optical spectroscopy (Sec.~\ref{sec:bloated_sdO}), and hot high-mass stripped stars can be characterized by their photometric UV excess, and with UV and/or visible spectroscopy \citep[e.g.][]{dro23, ram24}. 

Long time-series of optical spectroscopy has provided the known SB1 orbits of Be stars, thanks to dedicated monitoring of the short- intermediate- and long-time scale variations of this population \citep[e.g.][]{har00,ven11,naz22}. However, such efforts have not been made for Bn stars, which lack disks and thus are much less obviously variable. It should be noted that Bn stars do seem to pulsate in a similar manner as Be stars \citep{naz24}, and that at least some are post-interaction products (Sec.~\ref{sec:Be_sdO}). Dedicated monitoring of Bn stars is certainly warranted, both to conduct a binary census of the population, as well as to better understand their pulsational behavior and how this compares to that of Be stars. This may help to elucidate the processes by which Be stars create disks, while Bn stars do not.

Despite the relative dearth of good observational datasets suitable for the subtle binary analysis of Bn stars, the two lowest-luminosity stripped stars yet known belong to late-type Bn systems. In the case of KOI-81, this was rather serendipitous, with eclipses being found in Kepler photometry \citep{mat15}. The lack of a disk (which would otherwise increase complexity) in KOI-81 likely aided in the detection and characterization of this system. The companion to Regulus, on the other hand, was detected by sheer brute force through the analysis of an extensive set of high-quality optical spectra of this very bright star \citep{gie20}. This was made possible by the knowledge derived from its SB1 orbit \citep{gie08}. 

As demonstrated with KOI-81 \citep{mat15} and HD~92406 \citep{dea23}, finding Be/n stars with eclipses is in principle an excellent method for detecting binaries. However, owing primarily to the relatively long orbital periods and small stripped star sizes, an inclination angle very close to 90$^{\circ}$ is required to produce eclipses, and thus eclipsing systems are intrinsically rare. 

Recently, interferometry has been established as a powerful tool for detecting and characterizing Be/n+stripped binaries \citep{kle24}, and is the only technique capable of providing a model-independent astrometric orbit. 
For example, the stripped star in $\kappa$ Dra was first identified in interferometric measurements, and with knowledge of its orbital properties was recovered in IUE UV spectroscopy \citep{kle22}. 
This was the first confirmed Be+sdB system, indicating that not all stripped stars in Be binaries are hot sdO stars (with temperatures between about 40000 -- 50000 K). For systems with magnitudes below about 7, the CHARA interferometric array is capable of reaching contrast ratios as low as $\sim$0.3\% (e.g. in the H/K bands with the MIRC-X/MYSTIC instruments) in ideal conditions, corresponding to a magnitude difference of $\sim$6.4 \citep{gal15}, and capable of achieving an angular resolution of $\sim$0.5 mas. The VLTI interferometric facility offers similar capabilities for detecting high contrast close binaries. 
While interferometry is now a well established technique for discovering and characterizing Be+sdOB systems, very high contrast ratios as in Regulus \citep[with a $V$-band flux ratio of about 0.06\%,][]{gie20} remain out of reach.

There is no single `best' technique for discovering and characterizing Be/n binaries. Instead, the best results are achieved when combining different dataset and techniques which can depend on the specifics of the system(s) being analyzed. Nevertheless, UV spectroscopy has historically been the primary driver of this field, and remains an indispensable tool in directly detecting and characterizing stripped stars. The current lack of high-quality UV spectroscopy of a large sample of Be/n stars may be the main bottleneck in the advancement of this field.

\begin{figure*}[ht]
\centering
\includegraphics[width=0.99\hsize]{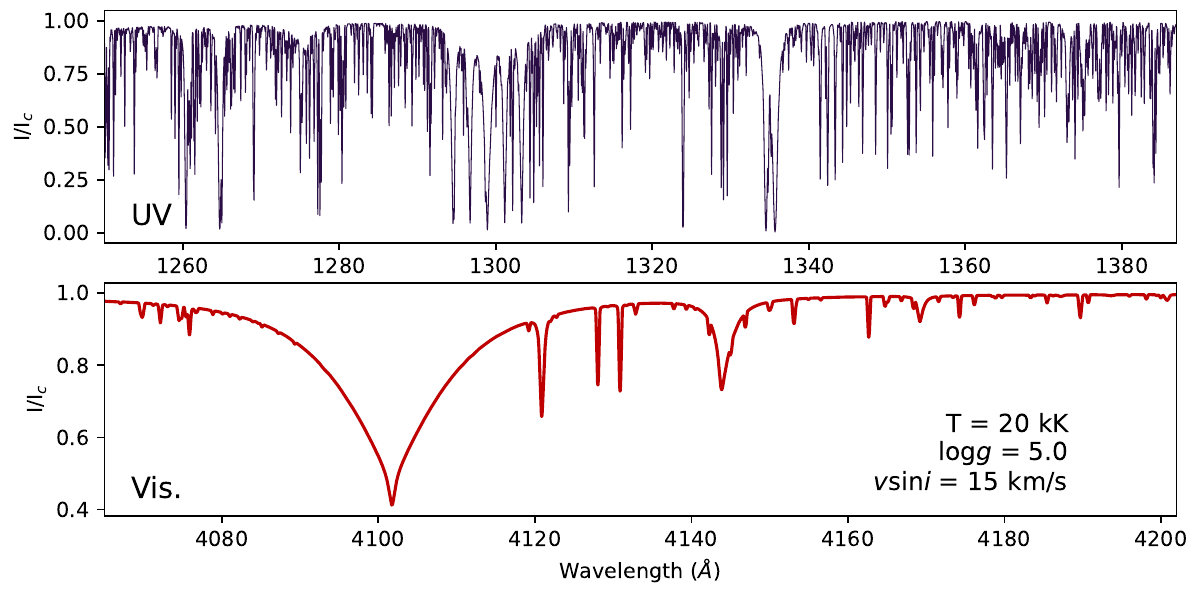}
\caption{Synthetic spectrum for an atmosphere with $T_{\rm eff} = 20000$ K, log$g$ = 5, and broadened to $v$sin$i$ = 15 km s$^{-1}$. The upper and lower panels show a 137 \AA\, portion of the UV and visible regions, respectively. }
\label{fig:UV_vs_Vis}
\end{figure*}

One advantage of working with UV spectroscopy compared to visible and IR wavelengths is that in all observed cases (excluding the bloated stripped stars) the stripped star is hotter than the Be/n star and so the contrast is most favorable at UV wavelengths. Another is that the UV regime is exceptionally rich with absorption lines for hot stars.
Figure~\ref{fig:UV_vs_Vis} compares the UV and visible regimes for a synthetic spectrum of an object qualitatively similar to an sdB or pre-WD star, with $T_{\rm eff} = 20000$ K, log\,$g$ = 5, and $v$sin$i$ = 15 km s$^{-1}$, using the line-by-line stellar atmosphere models \citep{shu04} as implemented in the PYSME\footnote{\url{https://github.com/MingjieJian/SME}} package. 
The multitude of features in the UV greatly enhance the power of cross correlation techniques that can be used to detect and characterize faint stripped stars in spectra dominated by a slightly cooler MS star. In practice, an additional advantage is that stripped stars generally have a very low $v$sin$i$ ($\lesssim$15 km s$^{-1}$) so that their lines are considerably sharper than those from rapidly rotating Be/n stars, assisting in efforts to disentangle the signatures of both components \citep{wan23}.

Historically, the limiting factor in detecting stripped stars with hot MS primaries with UV spectroscopy is the signal to noise ratio (SNR). In this sense, it is remarkable that the majority of known Be+sdO systems were first discovered in archival IUE spectra with typical SNR of $\sim$10 \citep{wan18}, and later confirmed with higher quality data \citep[e.g. from HST with SNR$\sim$30,][]{wan23}. However, from the archival IUE data, it was only possible to detect the most favorable such systems with relatively bright stripped stars. 
In this sense, the archival IUE UV spectroscopic database is likely nearly exhausted in its utility for discovering stripped stars, although it may continue to be useful for `pre-covery' of stripped stars following information derived from other methods \citep[as in e.g.][]{kle22}. HST currently provides the gold standard of UV spectroscopy, but tends to be used to confirm and better characterize Be/n+stripped systems after they have already been identified (perhaps as candidates) through other means \citep[e.g.][]{gie98, wan21, mat15}. 
Future high-resolution UV spectrographs will be needed to continue to investigate this field.  For example, the UVI and Pollux instruments \citep{mus18} considered for the Habitable Worlds Observatory (HWO) would be excellent facilities for this science topic, as well as smaller UV spectrographs or spectropolarimeters such as those planned on the SMEX candidate PolStar \citep{sco22} and the ESA mission candidate Arago \citep{mus22}.

\subsection{Filling in the gaps in the Be/n+stripped population} \label{sec:filling_the_gaps}

The current known population of Be/n+stripped systems may represent only the tip of the iceberg. 
Naturally, the brightest stars with the most favorable flux ratios are the easiest systems for detecting stripped companion. For example, the Be/n+sdOB stars plotted in Fig.~\ref{fig:HRD}, the median $V_{\rm mag}$ is 5.7 (omitting bloated and high-mass companions). There is an obvious dearth of sdB and pre-WD stars, with only four that are published, to the best of our knowledge. Considering also the large number of Bn stars that exist, which remain poorly observed (including by IUE, which led to the discovery of most of the Be+sdO systems), it now seems essential to properly explore the parameter space where Be/n+stripped binaries may be found.

Progress can be made towards better understanding this population on several fronts. Spectroscopic monitoring for RV motion, especially when combined with the often large archival databases for Be stars, 
is promising for discovering new binaries and determining orbital periods and eccentricities. Interferometry can continue to detect close binaries, or rule out certain types of companions. To better populate the current observational gaps, attention should especially be given to (non-emission line) Bn stars of all spectral types, and mid- to late-type Be stars.

For directly detecting stripped companions, Fig.~\ref{fig:BPASS_age_heatmap} provides a guide for the expected luminosity and temperature contrast for potential or candidate Be/n+stripped binaries. 
Although we note that comparing the BPASS population to the observed sample, stripped companions around a Be/n star of a given spectral type may be less luminous than the models predict. 
UV spectroscopy of moderate SNR ($\sim$30) has proven sufficient to detect sdO systems around early-type Be stars \citep[e.g.][]{wan21, wan23}, but higher SNR may be required for detecting helium fusing sdB stars with less favorable contrast ratios in the FUV. It should be noted that for some systems at certain orbital phases, the stripped companion is not detected, likely due to obscuration from circumstellar material \citep{pet16, wan21}. Stripped companions around Bn stars then should not suffer from this issue.

\begin{figure}[ht]
\centering
\includegraphics[width=0.99\hsize]{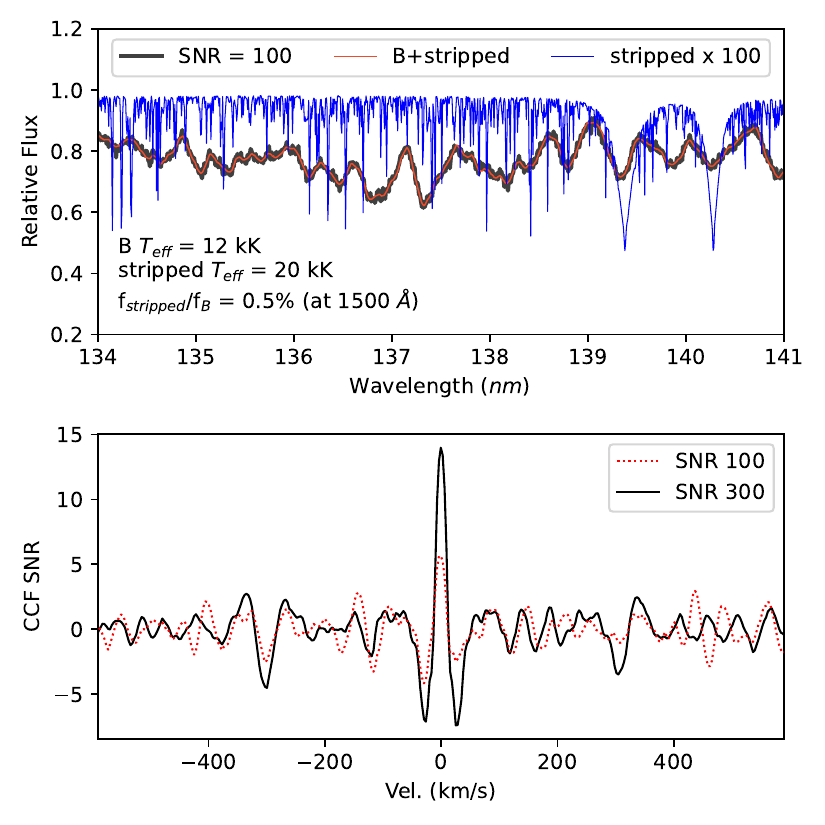}
\caption{Top: A portion of a synthetic UV spectrum emulating the Regulus system, with the primary having $v$sin$i$ = 300 km s$^{-1}$, $T_{\rm eff}$ = 12 kK, log$g$ = 3.5, and the stripped star having $v$sin$i$ = 15 km s$^{-1}$, $T_{\rm eff}$ = 20 kK, log$g$ = 5.0, with zero radial velocity offsets. Spectra from the two templates were co-added so that the stripped star contributes 0.5\% of the flux at 1500 \AA, consistent with the expected UV flux ratio based on the information in \citet{gie20}. The resolving power of the coadded spectrum was set to 20000. Bottom: The Cross Correlation Function (CCF) signal found after smoothing out the wide features (from the rapidly rotating primary) and using the stripped star spectrum as a template. The stripped star is marginally detected at a SNR of 100, and confidently detected at a SNR of 300. }
\label{fig:UV_Regulus}
\end{figure}

High-SNR far UV (FUV) spectroscopy is a viable pathway towards detecting low-mass stripped stars at high contrast ratios, especially for stripped stars unable to fuse helium. 
Figure~\ref{fig:UV_Regulus} shows a synthetic FUV spectrum that emulates the Regulus system \citep{gie20} by co-adding a spectrum of a main-sequence B star with $T_{\rm eff}$ = 12 kK and log$g$ = 3.5, and a stripped star with $T_{\rm eff}$ = 20 kK and log$g$ = 5.0, with a flux ratio of 0.5\% at 1500 $\AA$. For comparison, the sdB component of $\kappa$ Dra contributes $\sim$2.3\% of the FUV flux \citep{kle22}. The B-type spectrum was broadened to 300 km s$^{-1}$, and the stripped star to 15 km s$^{-1}$. The spectrum was degraded to a resolving power of $R=20000$, with noise added to simulate a SNR of 100 and 300. Employing cross correlation function (CCF) techniques, at a SNR of 100 the pre-WD in Regulus is marginally detectable, but at a SNR of 300, it is confidently detected. A UV spectroscopic survey at high SNR targeting the nearby population of rapid rotators would be a significant step forward in understanding the implications of binary interaction in high- and intermediate-mass stars.

\section{Discussion and Conclusions} \label{sec:conclusions}

Comparing the binary parameters of Be/n+stripped systems, SB1 Be stars, and the MS B+B binary population (Sec.~\ref{sec:compare_pops}) suggests the following. The SB1 systems are likely post-interactions binaries where the stripped star may be found along the helium MS and thus is expected to be detectable, or may be closer to a WD. 
For post-interaction binaries, the most likely nature of the companion depends on the spectral type of the Be/n star. For early-type primaries, the companion should either be an sdO star, or a (pre-)WD depending on the evolutionary stage. For mid-type primaries, the companion is most likely an sdB star. For later-type primaries, the companion may be a lower-mass sdB star fusing helium, or a (pre-)WD at low luminosity. However, the range of possible initial conditions that can lead to mass transfer and a Be/n+stripped configuration, the amount of time that has passed after the mass transfer stage, as well as parameters such as the degree to which mass transfer is conservative, prevent accurate predictions for the present-day configuration of a given specific system. Nevertheless, these trends seem to hold at the population level. 

Given the above, there is likely a cooler population of stripped stars, in the sdB and pre-WD regime that to date have hardly been observed. Direct detection of these sdB stars is possible with interferometry, UV spectroscopy, and possibly optical spectroscopy, while the pre-WDs can likely only be detected with high-SNR UV spectroscopy (Sec.~\ref{sec:filling_the_gaps}). Filling in these observational gaps is important to map the continuum of post-interaction evolutionary products, which for non-degenerate stripped stars ranges from the low-mass pre-WDs up to the high-mass quasi-WR stars (Sec.~\ref{sec:BPASS}). These higher-mass stripped stars may explode as SNe, producing NSs, which if they remain bound will become high-mass BeXRB systems (Sec.~\ref{sec:BeXRBs}) when the companion is a Be star, or can cause runaways (Sec.~\ref{sec:runaways}). The eventual fate of intermediate- and lower-mass stripped stars ($\lesssim$2 M$_{\odot}$) is most likely a WD. 

While this work focused primarily on the interacting systems that produce a rapidly-rotating B type star, it is worth considering a broader perspective. Binary interaction is even more frequent in O-type stars, but these spin down quickly. This makes using rapid rotation as a potential indicator of past interaction difficult. Relative to higher masses, binary interaction is understood to be less common in lower-mass stars. Nevertheless, some fraction of A- and F-type stars do experience binary interactions including mass transfer \citep[e.g.][]{kum24}, and there is also a rapidly rotating sub-population of A- and F-type stars \citep{roy07}. 
However, the Be phenomenon (where a star creates an outflowing gaseous decretion disk) is typically observationally limited to stars of spectral type $\sim$A2 and earlier, although may extend to slightly later stars \citep{anu21}. 
Stars spun up by binary interaction, however, are under no obligation to create an outflowing gaseous disk. The low fraction of Ae stars (and absolute lack of F-type disk makers) cannot be used as an argument against binary interaction producing rapid rotation in this lower mass regime. In this scenario, any stripped companions would surely be too low-mass to reach the He MS, and probably appear as (pre-)WDs.  

Recently, \citet{lec25} explored the properties of 16 Be+sdOB binaries to evaluate the efficiency of mass transfer during the interaction phase. Their results indicate that mass transfer is relatively efficient, such that most of the mass lost by the initial primary is gained by the accretor, strongly disfavoring a scenario where accretion is limited by a centrifugal barrier (i.e. when mass transfer only proceeds until the accretor is spun up to critical). Their results also appear to be in tension with mass transfer prescriptions limited by the thermal timescale of the accretor. The mass transfer prescription employed by BPASS is of this variety, with the accretion rate being limited by 
$\dot{M}_{2} \leq M_{2}/\tau_{\rm KH}$, where $M_{2}$ and $\tau_{\rm KH}$ are the mass and thermal timescale of the accretor, respectively \citep{eld17}. 

Although the qualitative agreement is good, our analysis also suggests tension between the observed Be/n+stripped population and the BPASS models. The average predicted post-interaction mass ratios, $q$, are about 30\% -- 60\% higher than in the observed sample (Figs.~\ref{fig:hist_of_q},~\ref{fig:BPASS_Q_hist}). Likewise, for a given region in the He MS, the corresponding mass gainer in BPASS is of lower mass (or $L$ and $T_{\rm eff}$) compared to what is observed (Fig.~\ref{fig:BPASS_age_heatmap}). 
In other words, for a given stripped star, the observations suggest a MS star with a $\sim$30\% -- 60\% higher mass than predicted by the BPASS models.  
These tensions may be be alleviated by incorporating a higher mass transfer efficiency in BPASS.  An increased efficiency of mass transfer in population synthesis models will result in a larger number of higher-mass stars, which in turn will produce a greater number of SNe. 

Accurate and precise measures of post-interaction stellar and binary properties and better statistics can then be used to improve parameters (such as how conservative mass transfer is) in binary stellar population synthesis codes. More accurate models, coupled with a better observational view of the Be/n+stripped population should allow for a reliable determination of what fraction of rapid rotators have been spun up by binary mass transfer. This has important implications for stellar astrophysics in general, and also at larger scales, such as the ionizing flux in clusters and star forming regions \citep{got19,got20}. 
Multiple aspects of the sort of binary interaction discussed in this paper contribute to a higher amount of ionizing UV flux than would be produced were both stars to evolve as single. The stripped star will be significantly hotter, albeit much smaller. 
The mass gainer will also have a higher effective temperature by virtue of its increased mass, while very rapid rotation will increase the polar temperature up to an additional $\sim$30\% \citep[e.g.][]{2019ApJS..243...10P}. If rotational mixing is efficient and the mass-accretion efficiency is not too low, then the main sequence lifetime may be extended by up to $\sim$30\% \citep{mae00}. 

It is worth mentioning the final fate of the massive post-interaction binaries.
Modern gravitational wave (GW) instruments, such as LIGO, have recently 
begun to explore black hole (BH) physics via observing the mergers of BH/NS+BH/NS pairs as they dynamically warp the fabric of spacetime \citep{abb17}. This is an exciting new frontier in probing nature at its most extreme. Such mergers are a possible outcome for the higher-mass post-interaction binaries, after both stars have exploded as a SN (or perhaps collapsed directly into a BH). The lower-mass post-interaction binaries may form very close WD+WD pairs before merging and igniting a SN explosion, which may also lead to GW signals detectable by future instruments such as LISA \citep{kor24}. It is therefore crucial to improve our understanding of the evolutionary pathways starting from the initial conditions of massive binaries, through their interaction and post-interaction stages in order to predict the final outcome of such systems, which will better enable us to trace back the history of observed GW events.

\bmhead{Acknowledgments}
The authors thank the anonymous referee whose thorough reading and comments served to improve the manuscript.
JLB is co-funded by the European Union (ERC, MAGNIFY, Project 101126182). Views and opinions expressed are, however, those of the authors only and do not necessarily reflect those of the European Union or the European Research Council. Neither the European Union nor the granting authority can be held responsible for them
CEJ and RGR acknowledge support from the Natural Sciences and Engineering Research Council of Canada. YN, a FNRS senior research associate, acknowledges support from FNRS and the Li\`ege university. AuD acknowledges support from NASA through Chandra Award number TM4-25001A issued by the Chandra X-ray Observatory 27 Center, which is operated by the Smithsonian Astrophysical Observatory for and on behalf of NASA under contract NAS8-03060.
This research has made use of the SIMBAD database operated at CDS, Strasbourg (France), and of NASA’s Astrophysics Data System (ADS).
This work made use of the PySME package for stellar spectral analysis.
We acknowledge the use of the Binary Population and Spectral Synthesis (BPASS) models, developed by Eldridge, Stanway, and colleagues.
This research made use of Astropy,\footnote{http://www.astropy.org} a community-developed core Python package for Astronomy \citep{astropy2013, astropy2018}.

\noindent

\bigskip

\onecolumn

\begin{appendices}
\clearpage

\section{Tables of confirmed binaries}\label{secA1}

This appendix contains tables describing the basic stellar and orbital properties of the post-interaction binaries gathered from the literature. Table~\ref{tbl:tbl_sdOB} includes the Be/n+sdOB systems, Tbl.~\ref{tbl:tbl_SB1} the SB1 Be binaries, and Tbl~\ref{tbl:tbl_bloated} the Be+bloated binaries. Table~\ref{tbl:tbl_highM} lists the high-mass stripped stars recently confirmed to be binaries. However, the stellar properties of the non-stripped companions in Tbl.~\ref{tbl:tbl_highM} are in general poorly determined. For the last nine entries in Tbl.~\ref{tbl:tbl_highM} the non-stripped star has not been directly observed, but RV motion of the stripped star indicates its binarity. In these cases, it is possible that the systems have evolved through a common envelope phase, rather than Roche lobe overflow, so that the non-stripped component is cannot at present be presumed to be rapidly rotating \citet{got23}.

\begin{landscape}
\begin{table*}
\caption{Basic properties of confirmed Be/n+sdOB binaries.}
\label{tbl:tbl_sdOB}
\centering
\setlength{\tabcolsep}{3pt}
\begin{tabular}{c c c c c c c c c c c c c}
\hline
HD & ID & ST & P$_{\rm orb}$ & e & K$_{\rm 1}$ & M$_{\rm 1}$ & log(T$_{\rm eff, 1}$) & log(L$_{\rm 1}$) & M$_{\rm 2}$   & log(T$_{\rm eff, 2}$) & log(L$_{\rm 2}$) & i \\
   &    &    & (d)           &   & (km s$^{-1}$)  & (M$_{\odot}$)  & (K)                      &  (L$_{\odot}$)      & (M$_{\odot}$) & (K)  & (L$_{\odot}$) & ($^{\circ}$) \\
\hline
200310 & 60 Cyg & B1Ve & 147.68$^{1,2,3}$ & 0.2$^{3}$ & 11.2$^{3}$ & 10.75$^{3}$ & 4.43$^{1}$ & 4.13$^{4}$ & 1.45$^{3}$ & 4.62$^{5}$ & 3.14$^{4,5,A}$ & 84.6$^{3}$\\
200120 & 59 Cyg &  B1.5Vnne  & 28.2$^{6,7}$ & 0.14$^{7}$ & 11.7$^{7}$ & 7.94$^{7}$ & 4.34$^{7}$ & 3.90$^{7}$ & 0.79$^{7}$ & 4.72$^{7}$ & 3.00$^{7}$ & 54.2$^{3}$\\
194335 & V2119 Cyg & B2IIIe & 63.146$^{2,3}$ & 0$^{3}$ & 12.2$^{3}$ & 8.2$^{3}$ & 4.41$^{8}$ & 3.98$^{8}$ & 1.35$^{3}$ & 4.64$^{9}$ & 2.94$^{9}$ & 49.5$^{3}$\\
55606 & HD 55606 & B0.5Vnnpe & 93.8$^{10}$ & 0$^{10}$ & 10.74$^{10}$ & 6.2$^{10}$ & 4.32$^{10}$ & 3.64$^{10}$ & 0.9$^{10}$ & 4.61$^{9}$ & 2.38$^{9}$ & 80$^{10}$\\
58978 & FY CMa &  B0.5IVe & 37.3$^{11}$ & 0$^{11}$ & 17.4$^{11}$ & 11.5$^{11}$ & 4.44$^{11}$ & 4.88$^{8}$ & 1.3$^{11}$ & 4.65$^{11}$ & 4.11$^{8,11,B}$ & 68$^{8}$\\
41335 & HR 2142 &  B1.5IV-Vnne  & 80.9$^{3,12}$ & 0$^{3}$ & 7$^{3}$ & 17.6$^{3}$ & 4.32$^{12}$ & 4.37$^{8}$ & 1.03$^{3}$ & 4.63$^{12}$ & -$^{}$ & 77.7$^{3}$\\
10516 & phi Per &  B1.5 V:e-sh  & 126.7$^{13}$ & 0$^{13}$ & 9.97$^{14}$ & 10$^{14}$ & 4.40$^{14}$ & 4.30$^{14}$ & 1.26$^{14}$ & 4.72$^{13}$ & 4.10$^{14}$ & 82$^{13}$\\
157042 &  iota Ara &  B2Vnne & 176.17$^{15}$ & 0$^{15}$ & 5.8$^{15}$ & 10.5$^{15}$ & 4.34$^{16}$ & 3.97$^{C}$ & 1.06$^{15}$ & 4.53$^{9}$ & 2.64$^{9}$ & 46$^{15}$\\
137387 & kap01 Aps &  B2Vnpe  & 192.1$^{15}$ & 0$^{15}$ & 9.09$^{15}$ & 11.8$^{15}$ & 4.38$^{8}$ & 4.01$^{8}$ & 1.6$^{15}$ & 4.60$^{9}$ & 2.65$^{9}$ & 60$^{15}$\\
113120 & LS Mus & B2IVne & 181.54$^{15}$ & 0$^{15}$ & 10.66$^{15}$ & 10.1$^{15}$ & 4.36$^{8}$ & 3.91$^{8}$ & 1.43$^{15}$ & 4.65$^{9}$ & 2.52$^{9}$ & 74$^{15}$\\
152478 & V846 Ara & B3Vnpe & 236.5$^{15}$ & 0$^{15}$ & 4.33$^{15}$ & 6.5$^{15}$ & 4.30$^{8}$ & 3.09$^{8}$ & 0.53$^{15}$ & 4.62$^{9}$ & 2.31$^{9}$ & 60$^{15}$\\
191610 & 28 Cyg & B3IVe & 359.26$^{2,3}$ & 0$^{3}$ & 5.4$^{3}$ & 6.26$^{3}$ & 4.31$^{8}$ & 3.83$^{8}$ & 0.76$^{3}$ & -$^{}$ & -$^{}$ & 118.7$^{3}$\\
161306 & EM* MWC 271 & B3/5Vnne & 99.3$^{3,17}$ & 0$^{3}$ & 5.6$^{3}$ & 6.02$^{3}$ & 4.21$^{C}$ & 2.89$^{C}$ & 0.784$^{3}$ & -$^{}$ & -$^{}$ & 34$^{3}$\\
109387 & Kap Dra & B6IIIe & 61.55$^{18}$ & 0$^{18}$ & 6.9$^{18}$ & 3.65$^{18}$ & 4.15$^{18}$ & 3.07$^{18}$ & 0.426$^{18}$ & 4.22$^{18}$ & 1.52$^{18}$ & 40$^{18}$\\
183537 & 7 Vul & B5Vne & 69.5$^{3}$ & 0$^{3}$ & 8.7$^{3}$ & 4.25$^{3}$ & 4.19$^{19}$ & 2.91$^{19}$ & 0.477$^{3}$ & -$^{}$ & -$^{}$ & $>$70$^{19}$\\
92406 & V658 Car & Bpsh & 32.185$^{20}$ & 0$^{20}$ & 12.2$^{20}$ & 4.45$^{20}$ & 4.15$^{20}$ & 2.51$^{20}$ & 0.53$^{20}$ & 4.34$^{20}$ & 2.26$^{20}$ & 88$^{20}$\\
87901 & Regulus & B8IVn & 40.102$^{21}$ & 0$^{21}$ & 7.58$^{21}$ & 3.7$^{21}$ & 4.07$^{22}$ & 2.38$^{23}$ & 0.31$^{21}$ & 4.30$^{21}$ & -0.30$^{21,D}$ & 86.3$^{24}$\\
- & KOI-81 & B8V & 23.876$^{25}$ & 0$^{25}$ & 6.74$^{25}$ & 2.92$^{25}$ & 4.07$^{25}$ & 2.13$^{25}$ & 0.19$^{25}$ & 4.30$^{25}$ & 0.14$^{25}$ & 89$^{25}$\\
60855 & V378 Pup & B3Ve & -$^{}$ & -$^{}$ & -$^{}$ & 5.4$^{C}$ & 4.30$^{9}$ & 3.99$^{26}$ & -$^{}$ & 4.62$^{9}$ & 2.85$^{9}$ & -$^{}$\\
43544 & HR 2249 &  B2/B3Ve & -$^{}$ & -$^{}$ & -$^{}$ & 8.5$^{9}$ & 4.33$^{27}$ & 3.55$^{26}$ & -$^{}$ & 4.58$^{9}$ & 2.70$^{9}$ & 46$^{8}$\\
51354 & QY Gem &  B3ne  & -$^{}$ & -$^{}$ & -$^{}$ & 5.4$^{C}$ & 4.30$^{9}$ & 3.49$^{26}$ & -$^{}$ & 4.64$^{9}$ & 2.85$^{9}$ & -$^{}$\\
29441 & V1150 Tau & B2.5Vne & -$^{}$ & -$^{}$ & -$^{}$ & 6.1$^{C}$ & 4.31$^{28}$ & 3.47$^{26}$ & -$^{}$ & 4.60$^{9}$ & 2.26$^{9}$ & -$^{}$\\
\hline
\end{tabular}
    \begin{tablenotes}
      \small
      \item In the column labels, the subscript ``1'' indicates the Be/n star, and ``2'' the stripped sdOB star. Note that when ranges of values are provided in a given paper, the average value is listed. References for these parameters are indicated by superscript numbers, and further notes with capital letters according to the following.
      1 = \citet{kou00}, 
      2 = \citet{kle22b},
      3 = \citet{kle24}, 
      4 = \citet{har00b}, 
      5 = \citet{wan17}, 
      6 = \citet{tar87}, 
      7 = \citet{pet13}, 
      8 = \citet{zor16}, 
      9 = \citet{wan21}, 
      10 = \citet{cho18}, 
      11 = \citet{pet08}, 
      12 = \citet{pet16}, 
      13 = \citet{mou15}, 
      14 = \citet{gie98}, 
      15 = \citet{wan23}, 
      16 = \citet{arc18}, 
      17 = \citet{kou14}, 
      18 = \citet{kle22}, 
      19 = \citet{har20}, 
      20 = \citet{dea23}, 
      21 = \citet{gie20}, 
      22 = \citet{bor23}, 
      23 = \citet{and12}, 
      24 = \citet{che11}, 
      25 = \citet{mat15}, 
      26 = \citet{naz22b}, 
      27 = \citet{lev06}, 
      28 = \citet{hoh10}, 
      A = Estimated from parameters in \citet{wan17}, considering luminosity from \citet{har00b}.
      B = Estimated from parameters in \citet{pet08} and \citet{zor16}. 
      C = Estimated from spectral type from the table of Eric Mamajek \url{https://www.pas.rochester.edu/~emamajek/EEM_dwarf_UBVIJHK_colors_Teff.txt}.
      D = Estimated from Fig.~8 in \citet{gie20}.
    \end{tablenotes}
\end{table*}
\end{landscape}

\begin{landscape}
\begin{table*}
\caption{Basic properties of confirmed SB1 Be binaries.}
\label{tbl:tbl_SB1}
\centering
\setlength{\tabcolsep}{3pt}
\begin{tabular}{c c c c c c c c c c c}
\hline
HD & ID & ST & P$_{\rm orb}$ & e & K$_{\rm 1}$ & M$_{\rm 1}$ & log(T$_{\rm eff, 1}$) & log(L$_{\rm 1}$) & M$_{\rm 2}$   & i \\
   &    &    & (d)           &   & (km s$^{-1}$)  & (M$_{\odot}$)  & (K)            &  (L$_{\odot}$)    &  (M$_{\odot}$) & ($^{\circ}$) \\
\hline
205637 & eps Cap & B3Vpe & 128.3$^{1}$ & 0$^{1}$ & 8.4$^{1}$ & 9.6$^{2}$ & 4.27$^{2}$ & 3.86$^{2}$ & 0.98$^{B}$ & 68$^{2}$\\
142926 & 4 Her & B9pe & 46.2$^{3}$ & 0$^{3}$ & 4.8$^{3}$ & 3.2$^{3}$ & 4.10$^{3}$ & 2.36$^{3}$ & 0.25$^{3}$ & 77$^{4}$\\
23862 & Pleione & B8Vne & 218$^{5}$ & 0.745$^{5}$ & 6.39$^{5}$ & 2.9$^{5}$ & 4.08$^{2}$ & 2.44$^{2}$ & 0.27$^{5}$ & 68$^{2}$\\
162732 & 88 Her & B8e & 86.7$^{6}$ & 0.15$^{6}$ & 8.91$^{6}$ & 2.9$^{2}$ & 4.15$^{2}$ & 2.62$^{2}$ & 0.41$^{6,B}$ & 62$^{2}$\\
183656 & V923 Aql & B7III & 214.716$^{7}$ & 0$^{7}$ & 3.02$^{7}$ & 6.2$^{2}$ & 4.15$^{2}$ & 2.95$^{2}$ & 0.45$^{7,C}$ & 88$^{2}$\\
184279 & V1294 Aql & B0.5IVe & 192.9$^{8}$ & 0$^{8}$ & 6.3$^{8}$ & 16.9$^{2}$ & 4.48$^{2}$ & 4.83$^{2}$ & 1.15$^{8}$ & 37$^{2}$\\
63462 & omi Pup & B1IV:nne & 28.9$^{9}$ & 0$^{9}$ & 10.3$^{9}$ & 13$^{9}$ & 4.55$^{2}$ & 4.43$^{2}$ & 0.7$^{9}$ & 67$^{2}$\\
212571 & pi Aqr & B1III-IVe & 84.1$^{10,11,12}$ & 0$^{11,12}$ & 8.1$^{11,12}$ & 10$^{12}$ & 4.38$^{10}$ & 3.87$^{12}$ & 0.8$^{12,D}$ & 56$^{4}$\\
5394 & gamma Cas & B0.5IVe & 203.523$^{13,14}$ & 0$^{13}$ & 4.3$^{13}$ & 13$^{15}$ & 4.46$^{15}$ & 5.19$^{2}$ & 0.98$^{13}$ & 45$^{16}$\\
37202 & zeta Tau & B1IVe-shell & 133$^{17,18,19}$ & 0$^{18}$ & 7.4$^{18}$ & 11$^{18}$ & 4.32$^{2}$ & 3.95$^{2}$ & 0.9$^{18}$ & 85$^{20}$\\
45995 & HD 45995 & B2Vnne & 103.1$^{19}$ & 0$^{19}$ & 6.7$^{19}$ & 10$^{19}$ & 4.40$^{2}$ & 3.79$^{2}$ & 1$^{19}$ & 45$^{2}$\\
183362 & V558 Lyr & B3Ve & 83.3$^{19}$ & 0$^{19}$ & 8.2$^{19}$ & 8$^{19}$ & 4.23$^{21}$ & 2.99$^{21}$ & 0.75$^{19}$ & -$^{}$\\
12882 & V782 Cas & B2.5III & 122$^{19}$ & 0$^{19}$ & 5.2$^{19}$ & 9$^{19}$ & 4.27$^{21}$ & 3.20$^{21}$ & 0.6$^{19}$ & -$^{}$\\
220058 & V810 Cas & B1 & 75.8$^{19}$ & 0$^{19}$ & 6.4$^{19}$ & 12.5$^{19}$ & 4.41$^{A}$ & 4.13$^{A}$ & 0.75$^{19}$ & -$^{}$\\
- & V2156 Cyg & B1.5V & 126.6$^{19}$ & 0$^{19}$ & 5.5$^{19}$ & 11$^{19}$ & 4.39$^{A}$ & 3.91$^{A}$ & 0.75$^{19}$ & -$^{}$\\
157832 & V750 Ara & B2Vne & 95.23$^{22}$ & 0$^{22}$ & 6.25$^{22}$ & 11$^{22}$ & 4.40$^{23}$ & 3.91$^{A}$ & 0.75$^{22}$ & -$^{}$\\
\hline
\end{tabular}
    \begin{tablenotes}
      \small
      \item References for these parameters are indicated by superscript numbers, and further notes with capital letters according to the following.
      1 = \citet{riv06},
      2 = \citet{zor16},
      3 = \citet{kou97},
      4 = \citet{sig23},
      5 = \citet{nem10},
      6 = \citet{due88},
      7 = \citet{wol21},
      8 = \citet{har22},
      9 = \citet{kou12},
      10 = \citet{bjo02},
      11 = \citet{naz19},
      12 = \citet{tsu23},
      13 = \citet{nem12},
      14 = \citet{rau22b},
      15 = \citet{har00},
      16 = \citet{kle24},
      17 = \citet{har84},
      18 = \citet{ruz09},
      19 = \citet{naz22},
      20 = \citet{car09},
      21 = \citet{pea13},
      22 = \citet{wan23},
      23 = \citet{lop11}.
      A = Estimated from spectral type from the table of Eric Mamajek \url{https://www.pas.rochester.edu/~emamajek/EEM_dwarf_UBVIJHK_colors_Teff.txt}.
      B = Estimated from mass function and inclination angle.
      C = Average value from a range of solutions.
      D = Estimated from their Fig. 8.
    \end{tablenotes}
\end{table*}
\end{landscape}

\begin{landscape}
\begin{table*}
\caption{Basic properties of confirmed Be binaries with stripped bloated companions.}
\label{tbl:tbl_bloated}
\centering
\setlength{\tabcolsep}{3pt}
\begin{tabular}{c c c c c c c c c c c c c c}
\hline
HD & ID & ST & P$_{\rm orb}$ & e & K$_{\rm 1}$    & K$_{\rm 2}$ &  M$_{\rm 1}$      & log(T$_{\rm eff, 1}$) & log(L$_{\rm 1}$) & M$_{\rm 2}$    & log(T$_{\rm eff, 2}$) & log(L$_{\rm 2}$)   & i \\
   &    &    & (d)           &   & (km s$^{-1}$)  & (km s$^{-1}$)  & (M$_{\odot}$)  & (K)                  &  (L$_{\odot}$)    &  (M$_{\odot}$) &    (K)     &   (L$_{\odot}$)     &  ($^{\circ}$) \\
\hline
- & LB-1 & B3V & 78.78$^{1}$ & 0$^{1}$ & 11.2$^{1}$ & 52.94$^{1}$ & 7.15$^{1}$ & 4.26$^{1}$ & 3.10$^{1}$ & 1.53$^{1}$ & 4.10$^{1}$ & 2.80$^{1}$ & 39$^{1}$\\
167128 & HR 6819 & B3IIIe & 40.3315$^{2,3,4}$ & 0$^{3}$ & 4$^{3}$ & 62.13$^{3}$ & 7$^{3}$ & 4.30$^{3}$ & 3.35$^{3}$ & 0.46$^{3}$ & 4.20$^{3}$ & 3.05$^{3}$ & 32$^{3}$\\
- & HIP 15429 & B5Ib & 221$^{5}$ & 0.52$^{5}$ & 5$^{5}$ & 75$^{5}$ & 7$^{5}$ & 4.23$^{5}$ & 4.43$^{5}$ & 0.7$^{5}$ & 4.13$^{5}$ & 4.36$^{5}$ & -$^{}$\\
190467 & V1362 Cyg & B3e & 56.82$^{6}$ & 0$^{6}$ & -$^{}$ & 82$^{6}$ & -$^{}$ & -$^{}$ & -$^{}$ & -$^{}$ & -$^{}$ & -$^{}$ & -$^{}$\\
698 & V742 Cas & B7:Ib-II(e) & 55.93$^{6}$ & 0$^{6}$ & -$^{}$ & 86.9$^{6}$ & -$^{}$ & -$^{}$ & -$^{}$ & -$^{}$ & -$^{}$ & -$^{}$ & -$^{}$\\
173219 & V447 Sct & B1:V:npe & 58.41$^{6}$ & 0$^{6}$ & -$^{}$ & 87.4$^{6}$ & -$^{}$ & -$^{}$ & -$^{}$ & -$^{}$ & -$^{}$ & -$^{}$ & -$^{}$\\
235679 & V2174 Cyg & BN2.5Ib:e & 225.33$^{6,7}$ & 0.075$^{6,7}$ & -$^{}$ & 61.1$^{6,7}$ & -$^{}$ & -$^{}$ & -$^{}$ & -$^{}$ & -$^{}$ & -$^{}$ & -$^{}$\\
44637 & MWC 139 & B3III:[n]e & $>$180$^{6}$ & -$^{-}$ & -$^{}$ & $>$30$^{6}$ & -$^{}$ & -$^{}$ & -$^{}$ & -$^{}$ & -$^{}$ & -$^{}$ & -$^{}$\\
\hline
\end{tabular}
    \begin{tablenotes}
      \small
      \item In the column labels, the subscript ``1'' indicates the Be/n star, and ``2'' the stripped star (which is lower mass but may be the more luminous component). The spectral types may be misleading, as such classifications may have been made prior to understanding the intricacies of these systems.  References for these parameters are indicated by superscript numbers.
      1 = \citet{she20},
      2 = \citet{riv20},
      3 = \citet{bod20b},
      4 = \citet{fro22},
      5 = \citet{mul25},
      6 = \citet{riv24},
      7 = \citet{bol01}.
    \end{tablenotes}
\end{table*}
\end{landscape}

\begin{landscape}
\begin{table*}
\caption{Basic properties of binaries with high-mass stripped stars.}
\label{tbl:tbl_highM}
\centering
\setlength{\tabcolsep}{3pt}
\begin{tabular}{c c c c c c c c c c c}
\hline
ID & ST & P$_{\rm orb}$ & e & K$_{\rm 1}$   &  M$_{\rm 1}$      & log(T$_{\rm eff, 1}$) & log(L$_{\rm 1}$) & M$_{\rm 2}$    & log(T$_{\rm eff, 2}$) & log(L$_{\rm 2}$)   \\
   &    & (d)           &   & (km s$^{-1}$) & (M$_{\odot}$)  & (K)                  &  (L$_{\odot}$)    &  (M$_{\odot}$) &    (K)     &   (L$_{\odot}$)    \\
\hline
2dFS 163$^{1}$ & O6.5Ib(f)+B0-1Ve & - & - & - & 10.5 & 4.41 & 4.18 & 3.96 & 4.57 & 4.75\\
2dFS 2553$^{1}$ & O4III(f)+B0-1Ve & 93.6 & - & - & 13.8 & 4.43 & 4.34 & 6.86 & 4.62 & 4.91\\
Sk -71 35$^{1}$ & B1Ia+O9Ve & 9.398 & - & - & 35.7 & 4.51 & 5.12 & 7.8 & 4.32 & 4.93\\
VFTS 291$^{2}$ & B5II-Ib+BV-IV & 108.03 & 0 & $<$15 & 13.2 & 4.44 & 4.41 & 2.2 & 4.12 & 3.96\\
1$^{3,4}$ &  & - & - & - & - & - & - & 8.45 & 4.95 & 5.09\\
2$^{3,4}$ &  & - & - & - & - & - & - & 3.31 & 4.80 & 4.12\\
3$^{3,4}$ &  & - & - & - & - & - & - & 3.38 & 4.85 & 4.15\\
4$^{3,4}$ &  & - & - & - & - & - & - & 3.04 & 4.83 & 4.01\\
5$^{3,4}$ &  & - & - & - & - & - & - & 4.06 & 4.82 & 4.34\\
6$^{3,4}$ &  & - & - & - & - & - & - & 3.74 & 4.86 & 4.24\\
7$^{3,4}$ &  & - & - & - & - & - & - & 2.91 & 4.79 & 3.95\\
8$^{3,4}$ &  & - & - & - & - & - & - & 2.14 & 4.76 & 3.55\\
16$^{3,4}$ &  & - & - & - & - & - & - & 1.63 & 4.52 & 3.20\\
\hline
\end{tabular}
    \begin{tablenotes}
      \small
      \item In the column labels, the subscript ``1'' indicates the Be/n star, and ``2'' the stripped star (which may be the more luminous component). The IDs for the last several columns are those used in \citet{dro23}. The first two entries in this table may have a classical Be star component \citep{ram24}. Sk -71 35 is possibly still undergoing mass transfer, so may be of a different nature than the rest of this sample. In most cases, details of the non-stripped companion are not yet known.   References for these parameters are indicated by superscript numbers.
      1 = \citet{ram24},
      2 = \citet{vil2023},
      3 = \citet{got23},
      4 = \citet{dro23}.
    \end{tablenotes}
\end{table*}
\end{landscape}




\end{appendices}


\bibliography{sn-bibliography}

\section*{Declarations}
RGR received funding from the Natural Sciences and Engineering Research Council of Canada (NSERC) Postgraduate Scholarship - Doctoral. CEJ received support from the NSERC Discovery Research Program. 
YN, a FNRS senior research associate, acknowledges support from FNRS and Univ. of Li\`ege.
AuD has received research support from NASA through Chandra Award number TM4-25001A issued by the Chandra X-ray Observatory 27 Center, which is operated by the Smithsonian Astrophysical Observatory for and on behalf of NASA under contract NAS8-03060. 
JLB received funding from the European Union (ERC, MAGNIFY, Project 101126182).
GJP received partial funding from the following NASA/STScI grants: HST-GO-17445.002-A, HST-GO-17095.003-A, and HST-GO-15659.002-A
The authors have no relevant financial or non-financial interests to disclose.

All authors contributed to the study conception and design. JLB collected and analyzed the observationally-determined information from the literature, with assistance from several authors (YN, KG, GP, RR, AR), and MS did the binary population synthesis modeling and subsequent analysis. All authors contributed text and expertise related to these topics, provided detailed comments which shaped the structure of the manuscript. All authors commented on previous versions of the manuscript, and have read and approved the final manuscript.

Ethics declaration: not applicable.

\end{document}